# Fossil and present-day stromatolite ooids contain a meteoritic polymer of glycine and iron.

[1*]Julie E M McGeoch, [2]Anton J Frommelt, [3]Robin L Owen, [4]Gianfelice Cinque, [5]Arthur McClelland, [6]David Lageson and [7]Malcolm W McGeoch

[1]Department of Molecular and Cellular Biology, Harvard University, 52 Oxford St., Cambridge MA  02138, USA & High Energy Physics Div, Smithsonian Astrophysical Observatory Center for Astrophysics Harvard & Smithsonian, 60 Garden St, Cambridge MA 02138, USA.
[2]LRL-CAT, Eli Lilly and Company, Advanced Photon Source, Argonne National Laboratory, 9700 S. Cass Avenue, Lemont, IL, 60439
[3,4]Diamond Light Source, Harwell Science and Innovation Campus, Didcot, OX11 0DE, UK.
[5]Center for Nanoscale Systems, Harvard University, 11 Oxford St, LISE G40, Cambridge, MA 02138, USA.
[6]Department of Earth Sciences, 226 Traphagen Hall, P.O. Box 173480 Montana State University, Bozeman, MT 59717.
[7]PLEX Corporation, 275 Martine St., Suite 100, Fall River, MA 02723, USA.
*Corresponding author. E-mail: Julie.mcgeoch@cfa.harvard.edu

**Abstract**
Hemoglycin, a space polymer of glycine and iron, has been identified in the carbonaceous chondritic meteorites Allende, Acfer 086, Kaba, Sutter's Mill and Orgueil. Its core form has a mass of 1494Da and is basically an antiparallel pair of polyglycine strands linked at each end by an iron atom. The polymer forms two- and three- dimensional lattices with an inter-vertex distance of 4.9nm. Here the extraction technique for meteorites is applied to a 2.1Gya fossil stromatolite to reveal the presence of hemoglycin by mass spectrometry. Intact ooids from a recent (3,000Ya) stromatolite exhibited the same visible hemoglycin fluorescence in response to x-rays as an intact crystal from the Orgueil meteorite. X-ray analysis confirmed the existence in ooids of an internal 3-dimensional lattice of 4.9nm inter-vertex spacing, matching the spacing of lattices in meteoritic crystals. FTIR measurements of acid-treated ooid and a Sutter's Mill meteoritic crystal both show the presence, via the splitting of the Amide I band, of an extended anti-parallel beta sheet structure. It seems probable that the copious in-fall of carbonaceous meteoritic material, from Archaean times onward, has left traces of hemoglycin in sedimentary carbonates and potentially has influenced ooid formation.

**Introduction**
Hemoglycin is a polymer of glycine, containing iron, that has been identified in extracts of five carbonaceous chondritic meteorites of primitive types that do not have extensive aqueous or thermal alteration. Having developed effective extraction and analysis techniques for these "stony" meteorites, we have applied them to first, a 2.1Gya fossil stromatolite, then to ooids of present-day stromatolite, to ask whether any trace of the compound could have reached the early Earth via in-fall and in some way interacted with biological systems. Two main properties that hemoglycin could contribute are its ability to form open lattices that can accrete materials, and the possibility that it could drive chemical processes via its newly discovered visible absorption band at 480nm. For background we present in Section S1 a short review of the current state of knowledge on hemoglycin's structure.

In this paper we employed four different experimental approaches in our comparison of stromatolite and meteoritic molecules: mass spectrometry; X-ray induced visible fluorescence; X-ray diffraction, and Fourier transform infrared (FTIR) absorbance. Each of them contributed to the conclusion that the meteoritic polymer hemoglycin was present in stromatolites, ancient and modern. New data emerged on the polymer in meteorites as well as in stromatolites, in particular relating to the secondary beta sheet structure adopted by hemoglycin in at least one meteoritic crystal.

Ooids, the primary mineral component of present-day stromatolites, are small ovoid particles (Figure 1) of mainly calcium carbonate that make up the oolitic sand that underlies and often overlays their formations. Fossil stromatolites have few intact ooids probably due to pressure over thousands to billions of years since their initial formation. Ooids in present day stromatolites contain calcium carbonate in the aragonite form, but over geological time there can be a transition within ooids to the calcite form, which is more energetically favored. The formation of ooids, and whether they form during stromatolite growth, or are merely pre-existing entities accreted into stromatolites, is still being researched. It is considered [1] that there is precipitation of calcium carbonate within stromatolite microbial mats via a matrix of extracellular polymeric substances, firstly in a calcium carbonate gel followed by production of nanospheres and then the growth of aragonite crystals guided by an organic matrix. On the other hand, Trower et al. [2] have documented independent ooid growth that is faster in a turbulent shallow water shoal (Turks and Caicos islands) than in a more static lagoon with more extensive biofilm colonization. There is extensive evidence for organics within biologically produced aragonite, coming from a) crystal anisotropy [3,4] and b) the presence in ooids of a blue fluorescence. In regard to b), Lin et al. [5] analyzed Holocene ooids (5,377±61Ya) from oolitic sand in the Western Qaidan Basin, Tibet. These were well-preserved, lacked microbial evidence, and contained 90-97% aragonite in fine crystals. The blue fluorescence under 365nm UV light was attributed to organic material within or between the crystals. In a study of present day stromatolites on Highborne Cay, Exuma, Bahamas, Paterson et al. [6] reported blue fluorescence of ooids under 405nm light, referring to it as "autofluorescence" to distinguish it from marker fluorescence. Elsewhere, Dravis et al. [7] reported strong fluorescence of ooids within a Pleistocene oolitic grainstone from West Caicos Island. The fluorescence was bright in well-preserved aragonite grains but freshwater diagenesis, that replaces aragonite with more stable calcite, apparently destroyed organic material and removed the fluorescence. These reports did not give fluorescence spectra in sufficient detail to allow comparison with the present x-ray induced visible fluorescence from ooids. A comparison of this fluorescence in meteorites [8] and ooids from a recent Shark Bay, Australia stromatolite, is presented below in Section 2 of this paper.

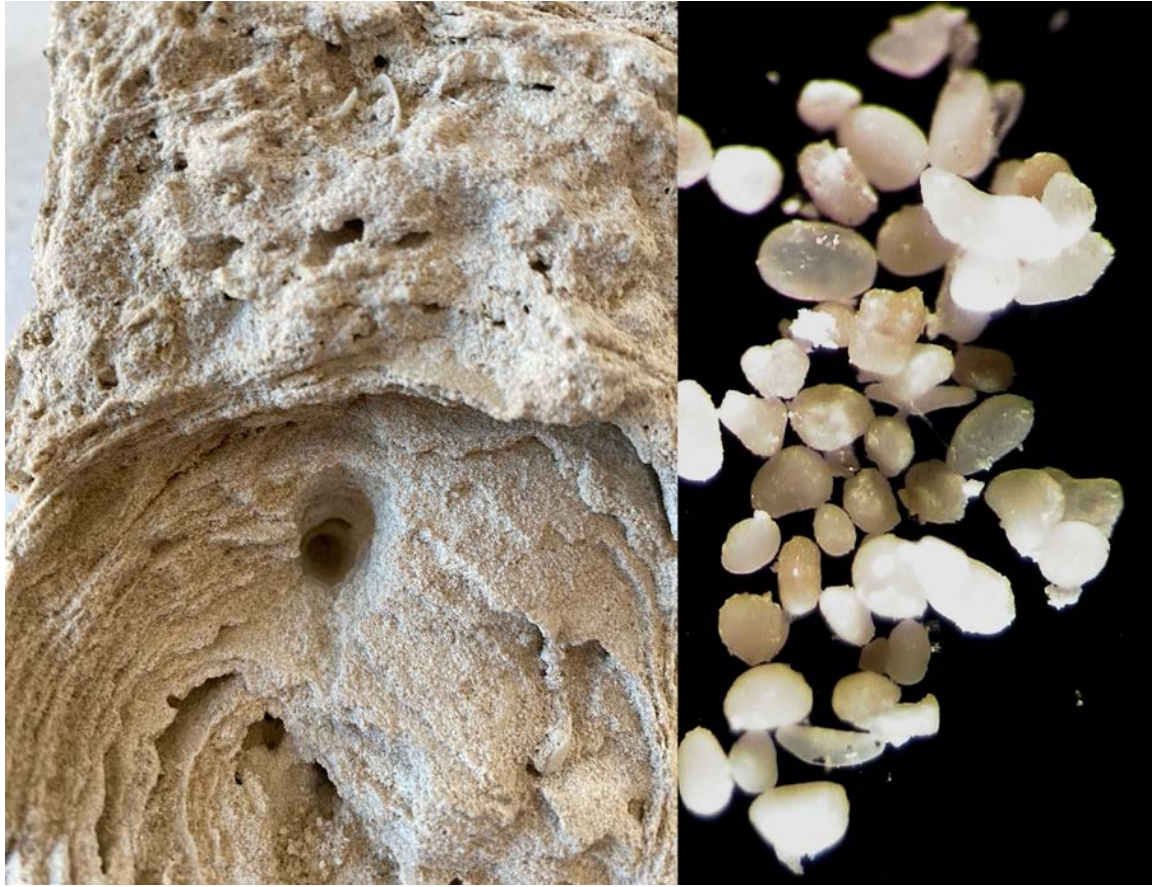

**Figure 1. Shark Bay stromatolite sample with drill hole (left) and Ooids (right) released from the sample by gentle etching. Ooid size range (39 yellow ooids): Major ooid axis 199±42μm; Waist diameter 164±39μm.**

In prior work on crystals of meteoritic hemoglycin we had observed two-dimensional lattices via the strong x-ray scattering of iron atoms at the lattice vertices, which are spaced by a hemoglycin polymer of length of 4.9nm. In [9] a floating 3-dimensional lattice was reported in the interphase region of the solvent extraction vial, and its structure was proposed to be the diamond 2H form. Subsequently, the tetrahedral angles of this form were seen in x-ray diffraction, but information was partial. As a matter of course the x-ray scattering of present day ooids was recorded, this time at wavelengths above and below the iron K-edge at 1.74 Angstroms so as to potentially know whether iron was involved in the organic internal lattice. At the longer 2.066 Angstrom wavelength used, where iron scattering was high and yet absorption was low, a dramatic series of high order diffraction rings was seen that did not appear at 0.979 Angstroms. Additionally, the anticipated aragonite and calcite rings at low "d" spacings were present in both cases. An analysis of this ooid data revealed a diamond 2H lattice of vertex spacing 4.9nm with a small axial lengthening above the ideal geometric lattice, presented below in Section 3 of this paper with more detail in supplementary Section S2.

The hemoglycin core unit [10,11] contains two antiparallel strands in a two-strand beta sheet. Infrared absorption data, discussed in Section 4 of this paper shows, in both a meteoritic crystal and an extract of a recent stromatolite ooid, the characteristic splitting of the Amide I poly-peptide band in the region of 6 microns that is only present in extended arrays of antiparallel beta sheets. This implies that in the meteorite crystal the hemoglycin core units are hydrogen-bonded edge-to-edge in an extended sheet (Figure S1.2), with the same being true for the structural protein in ooids, data presented in Section 4 of this paper.

In summary, mass spectrometry, x-ray induced visible fluorescence, x-ray diffraction and infrared absorption all point to the organic lattice of ooids being a molecule with the optical and spatial properties of meteoritic hemoglycin. The bulk of an ooid, filling this lattice, is a crystalline mixture of aragonite and calcite. In the discussion we assess meteoritic in-fall as the source of this lattice material.

**METHODS**
**Sample Sources:** Present-day stromatolites are supplied by Andrew Knoll of the Museum of Comparative Zoology and Organismic and Evolutionary Biology (OEB) Harvard.
Present-day stromatolite details are:
1. Shark-Bay Western Australia – collected by Elso Barghoorn 1971- estimated to be 2000-3000 years old.
2. **K-05 SS-1** from San Salvador Island Bahamas – collected by Andrew Knoll in 2005 –- A modern mineralized microbialite.

Fossil stromatolite details are:
1. 2.1Ga stromatolite No. 1 from Medicine Bow region, Wyoming – collected by David Lageson of Montana State University.
2. 2.1Ga stromatolite No. 2 from Medicine Bow region, Wyoming – collected by David Lageson of Montana State University.

**Mass Spectrometry**
The fossil stromatolite sample was collected from the Medicine Bow formations of the Wyoming craton [13]. The geological history of these rocks was relatively benign with an estimated maximum temperature underground of 300C, giving hope that molecular information could have survived to the present.
All experimental procedures were performed under clean laboratory conditions with operators wearing lab coats, hair cover and gloved hands as previously reported [3,12] (Figure 2) All chemicals were only used for these analyses and kept in separate laboratory areas. Micron particles of the fossil stromatolite were etched as previously described for meteorites and then Folch extracted [9,10,12] for up to 5 months at room temperature. The Orgueil meteorite sample being a total of 2 x 100mg samples from MNHM with a very loose topology typical of this meteorite, was not etched to micron particles but soft, small pieces of a few mg each were Folch extracted for 4 months.
Crystals of hemoglycin were picked up from the liquid interphase layer of the Folch extraction and pipetted into a watch glass on the stage of a zoom microscope under x25 magnification. Clean empty Hampton crystallography loops were used to pick up the 100-200µm crystals. The glycine rods of hemoglycin make the crystals slightly sticky allowing light adhesion to the loop. The Hampton loops were carefully placed into the inside of 500µl Eppendorf tubes containing 20µl of methanol. The loops were turned to make sure the crystals were delivered to methanol in the tubes and checked for this under the microscope. Several crystals were added to each Eppendorf tube. This technique of crystal transfer was used for both the fossil stromatolite and the Orgueil crystals. The fossil stromatolite crystals always had some adhering calcium carbonate particles. All stromatolite transfers were performed separate from those for Orgueil transfers to avoid any cross contamination. Each sample of crystals contained at least 5 separate crystals.

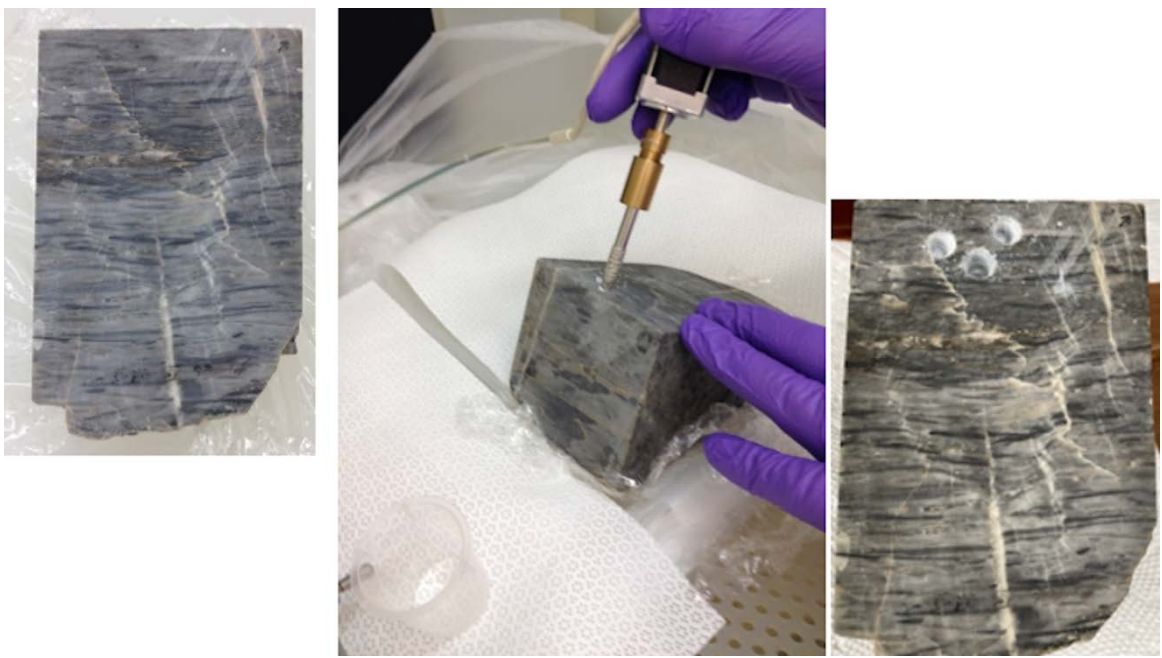

**Figure 2. Etch of the fossil stromatolite to produce micron particles for Folch extract in a clean room. Left, before etch; Middle - the etch using a stepper motor (no brushes to avoid metal contamination) with a vacuum brazed diamond drill bit (to avoid animal origin glue on drill bit); Right – 3 drill holes are visible. The micron particles are decanted by inversion of the sample over a glass container.**

A separate Orgueil Eppendorf tube was set up from a tube where hemoglycin crystals derived from unprocessed insoluble organic matter (IOM) had dried out and the crystals had adhered to the wall on the inside of the tube to an extensive degree (sample O3_1). Attempts were made using Folch solvents to solubilize these wall crystals which failed. It was decided as those crystals were potentially good because they resembled those that absorbed light [11], to simply leave the crystals of O3_1 as is, and rely on their solubilization via trifluoracetic acid at the point they are added to the SA and CHCA MALDI matrices, defined below.

After etching, two separate fossil stromatolite Folch extractions of 24 hours, were also set up as S3_1 for CHCA matrix and S3_2 for SA matrix. From these, 100µl aliquots from the interface layer were pipetted into Eppendorf tubes at 24 hours of extraction and allowed to reduce in volume to 20µl by loose cap room temperature evaporation. 2µl aliquots of each were used for the mass spectrometry analysis with paired CHCA and SA matrices.

In total there were 11 MALDI analyses performed, shown in Table 1. Throughout, the sample prefix is "S" for stromatolite and "O" for Orgueil.:

**Table 1. Detail of the 11 samples analyzed by MALDI mass spectrometry.**

| Sample source | Tube label | Matrix | Crystal origin |
|---|---|---|---|
| Fossil Stromatolite | S1_1 | CHCA | Mauve hexagonal crystals 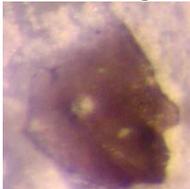 |
| Fossil Stromatolite | S1_2 | SA | Mauve hexagonal crystals 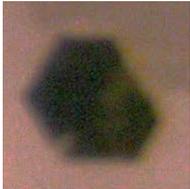 |
| Fossil Stromatolite | S2_1 | CHCA | Pale crystals 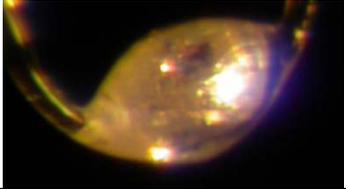 |
| Fossil Stromatolite | S2_2 | SA | Pale crystals |
| Orgueil meteorite | O1_1 | CHCA | Mauve hexagonal crystals 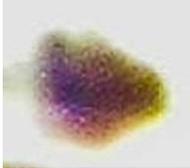 |
| Orgueil meteorite | O1_2 | SA | Mauve hexagonal crystals |
| Orgueil meteorite | O2_1 | CHCA | Pale crystals 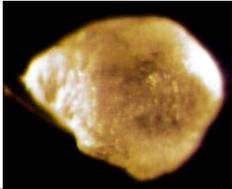 |
| Orgueil meteorite | O2_2 | SA | Pale crystals |
| Orgueil meteorite | O3_1 | CHCA | IOM Folch crystals |
| Fossil stromatolite | S3_1 | CHCA | Folch interphase |
| Fossil stromatolite | S3_2 | SA | Folch interphase |

Mass spectrometry was performed on a Bruker Ultraflextreme MALDI-TOF/TOF instrument. We used α-cyano-4-hydroxycinnamic acid (CHCA) and sinapinic acid (SA) matrix. Both were at 10 mg mL$^{-1}$ in 50% acetonitrile in water, 0.1% trifluoroacetic acid in water. Our resolution was of the order of 10,000 and we looked in the range m/z = 0–5,000, finding most peaks from 750-2000. A sample volume of 2µL was mixed with a matrix volume of 2µL, vortexed and left for one hour at room temperature. This one hour wait before pipetting 1µl quantities onto the MALDI plate is essential as it takes that long to partly solubilize hemoglycin in the matrix solvents.

**X-ray induced visible fluorescence of ooids**
The Shark Bay recent stromatolite (Figure 1) was provided by Dr. Andrew Knoll. Drilling as above (into fossil stromatolite) did not produce micron particles because the material was friable, breaking up into 200-500 micron scale fragments, accompanied by intact ooids (Figure 1).
All experimental procedures were performed under clean laboratory conditions with operators wearing lab coats, hair cover and gloved hands as previously reported [9,10,12]. All chemicals were only used for these analyses and kept in separate laboratory areas.
Ooids were placed in a watch glass and manipulated under X25 magnification. Ethyl cyanoacrylate glue (1/5$^{th}$ the volume of the ooid) was applied to a Hampton crystallography loop and the ooid attached to the loop by gently touching the glue to the crystal. After the glue solvents had evaporated (4 hours at 18C) the ooid post and loop assembly was capped, attached to a pad to prevent vibration and was send by FEDEX to Diamond Light Source. The internal structure of an ooid was obtained by treating ooids in a watch glass with 5% acetic acid. Within 5-10 minutes the internal vesicular structure was revealed (Figure 3).

X-ray induced visible fluorescence data were collected at a wavelength of 1.000 Angstroms (12.40 keV) and with a beam size of 50µm × 50µm. The flux of the unattenuated beam was 8 × 10$^{12}$ ph/s and fluorescence data was collected with the beam attenuations of 2%, 5%, 10%, 20%, 40%, 60%, 80%, 100%. UV-Visible data were collected *in situ* using off axis reflective objectives and were recorded over the wavelength range 250-800nm using an Andor shamrock 303i spectrograph and CCD detector.

**X-ray structural analysis**
Two X-ray wavelengths, 0.979 Angstroms and 2.066 Angstroms, were used on the APS beamline 31-ID-D. Data were recorded using a Pilatus3 S 6M detector with a detector distance of 190 mm except where noted, 1degree oscillations, and exposure times of 0.5 seconds. On Diamond beamline I24 diffraction data were recorded using a Pilatus3 6M detector with a detector distance of 300mm using 0.1deg oscillation per frame, exposure times of 10ms and beam attenuated by 50%.

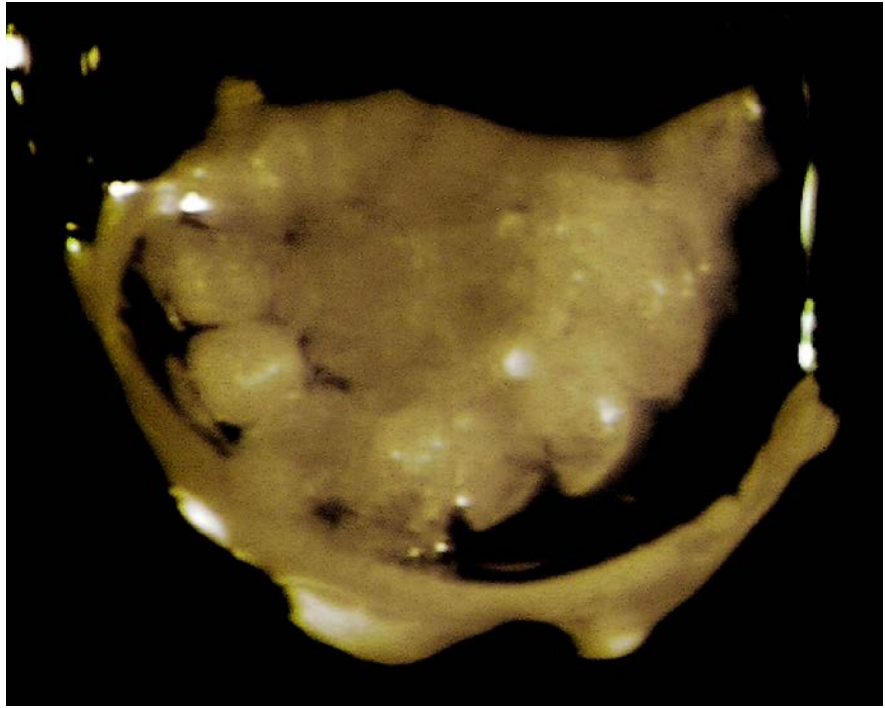

**Figure 3. 5% acetic acid treated ooid from present-day stromatolite, Shark Bay, Australia. The ooid (sample LS2) is on a crystallography loop for x-ray diffraction analysis. The individual vesicles revealed via the acid treatment are 20μm diameter.**

**Infrared absorption spectra**
**a) Meteorite sample analyzed on beamline B22, Diamond Light Source.**
A Sutter's Mill crystal (SM2) previously analyzed for structure and visible light absorption [11] was transferred from beamline I24 to B22 at Diamond Light Source for multimode infrared imaging and microspectroscopy. This crystal was kept on its loop for the IR absorption at B22 supported by modelling clay. The crystal was first scanned on its entire area for IR absorption and the amide I and II bands were concentrated in approximately 40 X 20 microns in an ordered crystalline area near the center of the loop. Data from this area was collected in detail and compared with surrounding areas. This indicated which bands correlated with amide and as opposed to belonging to non-amide content.
**b) Stromatolite Ooid sample analyzed at Harvard Center for Nanoscale Systems (CNS MET-15 Bruker FT-IR Lumos 1 Microscope).**
Present-day stromatolite ooids were placed under X25 magnification on glass slides containing carbon tape to stably secure each ooid. X-ray data had revealed these ooids contained principally the aragonite form of calcium carbonate. It was found this calcium carbonate content dominated the IR absorption spectrum at wavelengths 6.9, 11.7 and 14microns. Mild acetic acid treatment was applied to reduce this calcium carbonate absorption signal. Ooids secured by carbon tape on glass slide were placed in glass stain containers which contained 5% acetic acid or distilled water. First, for times between 5 minutes to 1hour the slides were in the acetic acid followed by washes in 4 separate containers containing distilled water. At the end of this acid treatment regime the slides were air dried and analyzed for IR absorption. It was found that 20 minutes was the ideal time to remove enough calcium carbonate from the ooids to produce a clear IR absorption related to extended antiparallel beta sheets (Amide I at 6μm). At 20minutes an amorphous calcium carbonate absorption was also present.

# RESULTS

## Section 1. Mass spectrometry on fossil stromatolite compared to Orgueil meteorite.

### 1.1 Approach
Our decision to employ MALDI mass spectrometry, in this and all our prior analysis was impelled by:
1. Unknown phase mixtures can be handled.
2. A useful degree of laser fragmentation contributes to the structural analysis.
3. There are two stand-out matrix molecules, CHCA and SA, that have reliable, yet very different protonation rates [14]. When their results coincide, uncertainty is removed.
4. Even when collections of small crystals are studied, as in this work, it is not necessary to grind the crystals as partial solvation is achieved in MALDI matrix solutions after 1 hour at room temperature.

We came into the measurements knowing that the crystals likely contained hemoglycin because:
1. They all derive from the interphase of the relevant Folch extractions, where hemoglycin was the dominant chemical [10].
2. X-ray diffraction has not been possible on these crystals due to their small size. However, on several larger crystals from both Orgueil and Sutter's Mill there were diffraction rings that formed a family related to specific iron atom spaces in the hemoglcyin polymer junctions.

To anticipate the results, we have obtained mass spectrometry confirmation in both Orgueil and fossil stromatolite that the 1494Da hemoglcyin core unit [10] comprises essentially 100% of the crystalline material that is able to be solubilized.

### 1.2 Basic observations from Mass spectrometry.
A very consistent MS pattern was seen in all samples, indicating that the crystal preparation process had universally selected the same molecule, whether from Orgueil or Stromatolite. In every sample the MALDI spectrum contained only two main peaks, one at 1494 m/z, corresponding to the hemoglycin "core unit" [10]and the second at 760 m/z, which was typically five times higher in summed ion count and was the exclusive fragment of 1494 m/z observed in the present work. The 760 m/z fragment comprises a single polyglycine strand from the antiparallel pair that comprise the central body of hemoglycin. It does not show $^{54}$Fe side peaks [10] beside the "monoisotopic" peak, and hence the fragment does not carry iron.

Figure 4 shows for Orgueil crystal sample (O1_2, run1, SA) the 760 m/z peak system on the left, and the 1494 m/z system on the right. They were each fitted to a global $^2$H enhancement, giving via 760 m/z analysis δ $^2$H = 54,000 ± 3,000 per mil, and via the 1494 m/z analysis δ $^2$H = 51,000 ± 3,000 per mil.

In summary, the strong polymer rod interconnections of hemoglycin rendered MS analysis for the 1494 m/z subunit itself fairly difficult. This was an inevitable side-effect of having a strong space polymer. The difficulty persisted in spite of mixing in MALDI solvent for 1hour, or 72hrs (run 2). Earlier it had been established that 1 hour in the matrix with 50% acetonitrile and 0.1%TFA was necessary to get the polymer solvated. Immediate MALDI approximately 15 minutes after applying to the MALDI plate, yielded no peaks at all with the polymer not leaving the crystals for the matrix.

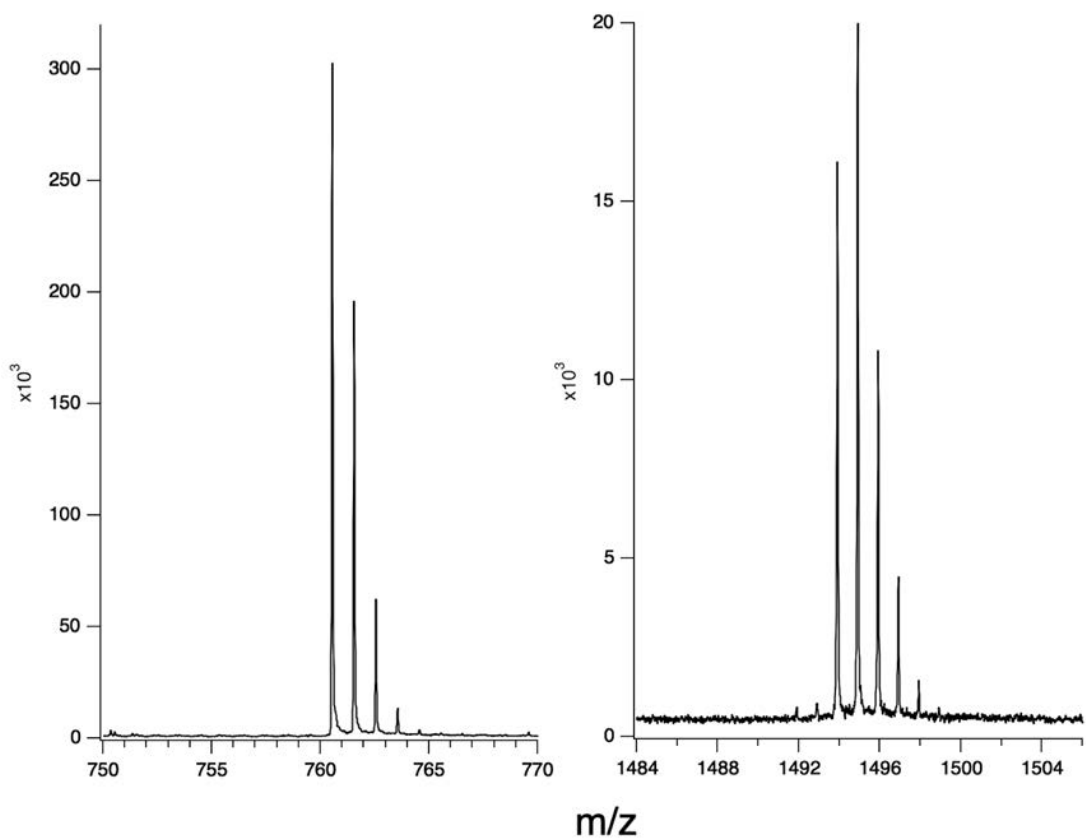

**Figure 4. Data from sample O1_2, with sinapinic acid matrix. On the left is the 760 m/z fragment that is the sole product from the break-up of the 1494 molecules. On the right, the 1494 m/z peak system.**

### 1.3 Isotope analysis
The first identification of hemoglycin in meteorites via MALDI MS [10] showed that it carried heavy isotopes ratios such as $^2$H/H, $^{13}$C/$^{12}$C, $^{15}$N/$^{14}$N, etc. at levels far above terrestrial standards, at least in the cases of $^2$H and $^{15}$N. An isotope fitting routine was written [10] with no internal approximations, in which trial values of the isotope ratios are input and the output is compared to experimental MALDI peak strengths. Figure 5, curve A, shows a stromatolite hemoglycin molecular peak complex at m/z 1494 (sample S2_1 (CHCA)). The highest isotopologue occurs at the n = +1 location relative to the n = (0) "monoisotopic" peak. Curve B in Figure 5 is the calculated complex for a "global" (i.e. equivalent $^2$H) enhancement of 52,000 per mil, but it is expected that in practice there would be a significant $^{15}$N component, as seen in Acfer 086 [10,15]. This would reduce the actual $^2$H/H ratio in a predictable way. "Global $^2$H" fitting to the stronger run2 data set, from both the 1494 and 760 components, is summarized in Table 2. In contrast, a simulation with wholly terrestrial isotope levels gives the very different curve (Figure 5 C), in which the highest peak is the (0) isotopologue.

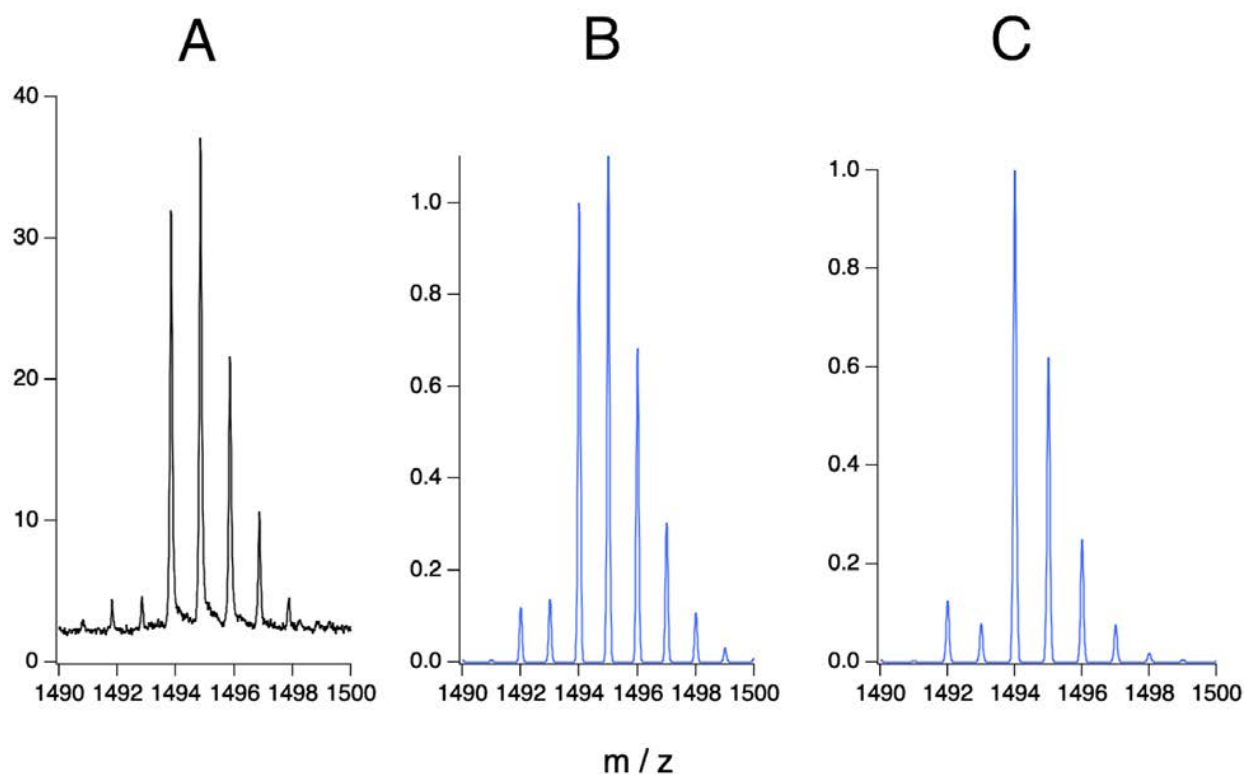

**Figure 5. Vertical axis is intensity and horizontal axis m/z. A: sample S2_1 peak complex at 1494 m/z. B: variation of $^2$H to fit the isotopologue intensities in curve A (the fit required $\delta^2$H = 52,000 per mil). C: the same molecule simulated at terrestrial (Vienna) isotope values.**

The following isotope ratios (IAEA, Vienna 1995 [16]) are taken as terrestrial standards:
VSMOW water         $R_H$ = $^2$H/$^1$H = 155.76 ± 0.05 x 10$^{-6}$
VSMOW water         $R_O$ = $^{18}$O/$^{16}$O = 2,005.20 ± 0.45 x 10$^{-6}$
V-PDB               $R_C$ = $^{13}$C/$^{12}$C = 11,237.2 x 10$^{-6}$
Atmospheric Nitrogen $R_N$ = $^{15}$N/$^{14}$N = 3,612 ± 7 x 10$^{-6}$

**TABLE 2. Isotope analysis for run2 data set. Enhancements in parts per mil.**

|  |  | Global $^2$H 760 m/z | Global $^2$H 1494 m/z |
|---|---|---|---|
| Orgueil | O1_1 (CHCA) | xx | 55,000 ± 2,000 |
|  | O1_2 (SA) | xx | 54,000 ± 3,000 |
|  | O2_1 (CHCA) | 50,000 ± 2,000 | xx |
|  | O2_2 (SA) | 51,000 ± 2,000 | xx |
|  | O3_1 (CHCA) | 51,000 ± 2000 | 53,000 ± 3,000 |
|  |  | Orgueil Ave. 52,333  σ = 1,795  n = 6 | |
| Stromatolite | S1_1 (CHCA) | xx | 53,000 ± 3,000 |
|  | S1_2 (SA) | xx | 53,000 ± 3,000 |
|  | S2_1 (CHCA) | xx | 52,000 ± 3,000 |
|  | S2_2 (SA) | xx | 54,000 ± 2,000 |
|  | S3_1 (CHCA) | xx | 51,000 ± 2,000 |
|  | S3_2 (SA) | xx | xx |
|  |  | Stromatolite Ave. 52,600  σ = 1,019  n = 5 | |
| "xx" represents saturated signal or too low signal relative to noise | | | |

## Section 2.  Visible ooid fluorescence induced by x-rays

Modern ooids were subject to x ray diffraction on both the Advanced Photon Source, Argonne National Laboratory (APS) and the Diamond Light Source (Diamond), with diffraction results reported in Section 3. An ooid (sample LS2) from Hamlin Pool, Shark Bay, Western Australia of estimated age 2,000 – 3,000 years, that had been treated with acetic acid as described in the Methods section to partially remove calcium carbonate, was the subject of x ray analysis at Diamond Light Source.
Sample LS2 gave strong x ray induced fluorescence in a 1.000 Angstrom beam, under cryo-flow cooling (Figure 6). The fluorescence was associated with low temperatures (100K) and was absent at 300K. The fluorescence peaked at 480nm and carried the same 465nm absorption "dip" seen previously in x ray induced fluorescence [8] in a crystal from the Orgueil meteorite, the data for both cases being shown in Figure 6.

The fluorescence was analyzed as before [8] using five Gaussian components to obtain an exact comparison. Across the range of wavelengths the same five components appeared, with relatively minor changes to the center wavelength of any one component. The intensity distribution was different, however, with much reduced representation of the peaks at 505nm and 565nm relative to the corresponding ones in the Orgueil data at 489nm and 551nm. These distributions are compared in Table 3. Intensities are multiplied by widths and normalized to the 489nm Orgueil peak, which is assigned the value 100.  In this ooid sample the dominant emission is still associated with a peak close to 480nm. However, the next strongest emissions are at 407nm and 477nm, rather than at 551 (or 565) nm. All of these emissions are linked to the interaction of iron with glycine residues [8,11].

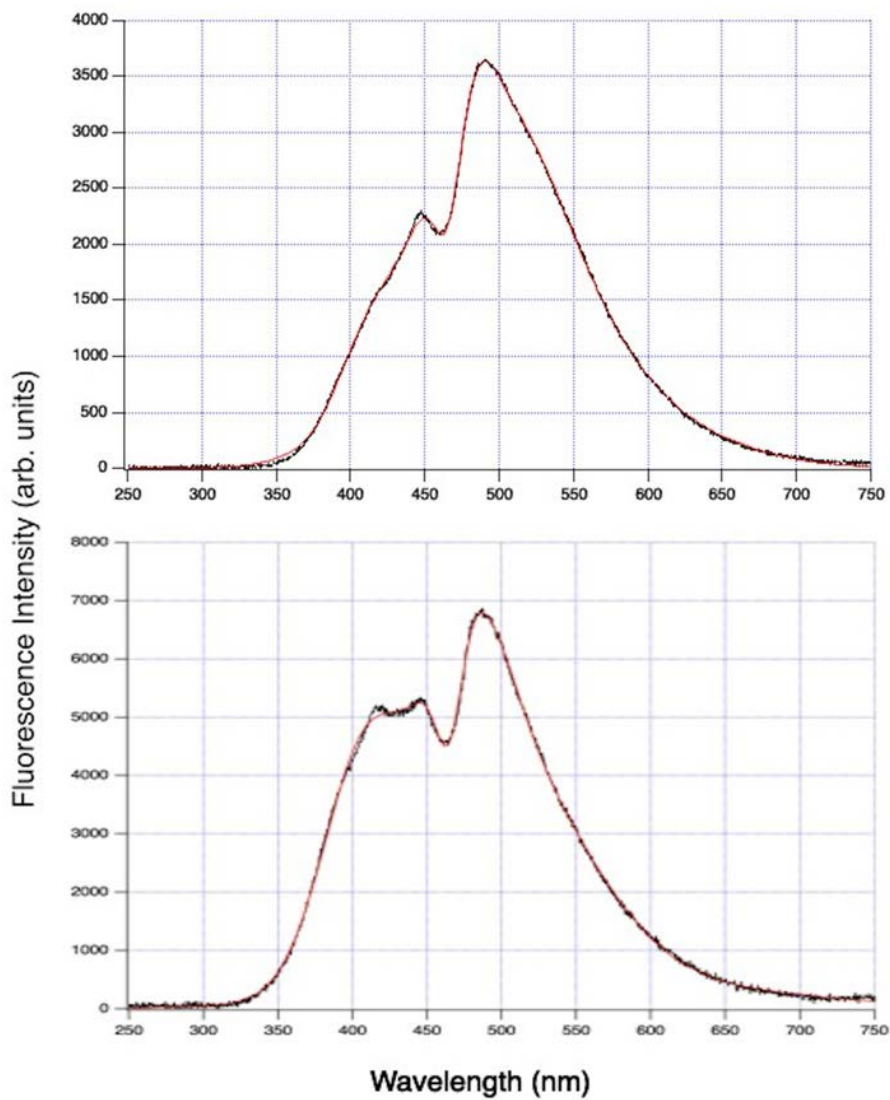

**FIGURE 6. X-ray induced blue-green fluorescence from (Top) Orgueil meteorite crystal; (Bottom) Ooid from present day Shark Bay Stromatolite. Experimental data black, fit curves in red.**

The ultraviolet and visible light absorbance of the sample (not shown) is featureless, not showing the characteristic 480nm absorbance seen in the Orgueil crystal [8] that is indicative of the chiral 480nm absorbance of hemoglycin [11]. This suggests that in the ooid sample there is probably a low population of 'R' chirality hydroxy-glycine residues bonded to Fe at their C-termini. More specifically, reviewing the visible absorptions calculated in Table 2 of [11] we deduce that there appear to be low populations of {0,R}, {S,0}, {S,R}, {R,0},

{R,S}, {R,R} combinations of {N-terminus, C terminus} chiralities of hydroxyglycine, with zero representing plain glycine. It is possible that the acetic acid treatment of this sample caused chemical reduction of most of the hydroxyl groups, leading to un-observable 480nm absorbance. Without the full complement of hydroxyl groups in the ooid, the fluorescence routes corresponding to the main Orgueil bands at 489nm and 551nm involving hydroxyglycine would be less able to function, consistent with observation.

Based on these compelling similarities in the fluorescence of ooids and the previously characterized Orgueil crystal, it is very likely that a similar iron-glycine chemical compound is in each sample. Observation of the sharp 465nm absorption again in ooids suggests that there could be a "caged" iron atom at junctions of the ooid hemoglycin lattice, as previously proposed for the Orgueil crystal [8].

**Table 3. Peak x ray induced visible fluorescence wavelengths, half widths at 1/e intensity, and relative integrated strengths, for an Orgueil crystal [8] and ooid from stromatolite. Errors were generated in least squares fitting of averages of 3 data traces taken at 100% beam intensity. Peak 4 is the absorption dip, represented by a negative intensity.**

| Peak | 1 | 2 | 3 | 4 | 5 |
|---|---|---|---|---|---|
| ORGUEIL $\lambda$(nm) | 408.5 ±0.4 | 489.0 ±0.3 | 551.2 ±9.7 | 465.1 ±0.2 | 488.5 ±0.5 |
| 1/e half width (nm) | 27 | 72 | 100 | 12.0 | 15.3 |
| Peak intensity | 408 ±16 | 2,826 ±135 | 714 ±98 | -759 ±15 | 339 ±10 |
| Integrated strength normalized | 5.4 | 100 | 35 | -4.5 | 2.5 |
|  |  |  |  |  |  |
| OOID $\lambda$(nm) | 407.3 ±0.9 | 504.9 ±7.4 | 565 ±15 | 464.8 ±0.2 | 477.4 ±0.8 |
| 1/e half width (nm) | 39 | 79 | 166 | 12.7 | 37 |
| Peak intensity | 389 ±33 | 357 ±33 | 59 ±5 | -202 ±5 | 353 ±42 |
| Integrated strength normalized | 54 | 100 | 36 | -9 | 46 |

# Section 3. X ray derivation of the three-dimensional hemoglycin lattice in ooids

Intact ooids from the Shark Bay stromatolite sample were studied for lattice structure at APS using x ray wavelengths of 0.979 Angstroms and 2.066 Angstroms which straddled the 1.74 Angstrom K absorption edge of iron. At each wavelength there were diffraction rings between principally 1.61 Angstroms and 3.85 Angstroms in a superposition of calcite and aragonite powder pattern rings (data in S2, Table S2.1 and Figure S2.2, part B). However, at 2.066 Angstroms, where Fe absorption is low, and not at 0.979 Angstroms, there was an intense and striking new set of rings (Figure 7 and Figures S2.1, S2.2) at nominal first order spacings of between 4.808 and 11.540 Angstroms, summarized in Table 4.

**Table 4. Ooid diffraction rings in first order (left hand column). Higher order fits (top row) listed as diffraction order in bold with percentage mis-match.**

| Angstroms | 49.0 | 81.65 | 92.05 | 112.75 | 119.9 | 126.25 |
|---|---|---|---|---|---|---|
| 4.80(8) |  | **17**, 0.1% |  |  | **25**, 0.2% |  |
| 5.18(5) |  |  |  |  |  |  |
| 5.24(3) |  |  | **17**, 0.1% |  |  | **24**, 0.3% |
| 5.44(5) | **9**, 0% | **15**, 0.2% |  |  | **22**, 0.1% |  |
| 5.63(3) |  |  |  | **20**, 0.1% |  |  |
| 5.73(5) |  |  |  |  | **21**, 0.4% | **22**, 0.1% |
| 5.94(3) |  |  | **15**, 0.1% | **19**, 0.1% |  |  |
| 6.28(8) |  | **13**, 0.4% |  | **18**, 0.4% | **19**, 0.3% | **20**, 0.4% |
| 6.57(8) |  |  |  |  |  |  |
| 6.84(5) |  |  | **13**, 0.1% |  |  |  |
| 7.02(2) | **7**, 0.3% |  |  | **16**, 0.3% | **17**, 0.4% | **18**, 0.1% |
| 7.12(8) |  |  |  |  |  |  |
| 7.48(5) |  |  |  | **15**, 0.4% | **16**, 0.1% |  |
| 8.12(7) | **6**, 0.5% | **10**, 0.1% | **11**, 0.3% |  |  |  |
| 9.05(3) |  | **9**, 0.2% |  |  |  | **14**, 0.3% |
| 9.82(3) | **5**, 0.2% |  |  |  |  |  |
| 10.21(7) |  | **8**, 0.1% |  | **11**, 0.3% |  |  |
| 11.54(0) |  |  |  |  |  |  |

The left hand column of Table 4 contains the set of 18 rings that represented larger "d" spacing than 4 Angstroms. These rings did not match either the calcite or aragonite values but were reminiscent of the ladders of high order diffraction previously seen in hemoglycin lattices [9, 11]. In [39] the ladder contained orders 2 through 5, with a fitted first order lattice parameter of 48.38 ± 0.2 Angstroms. In [11] the ladder contained orders 2 through 12, with a fitted first order parameter of 49.03 ± 0.18 Angstroms. Quick inspection of the present 18 ring set yielded a fit to 49.0 Angstroms in 5$^{th}$ 6$^{th}$ 7$^{th}$ and 9$^{th}$ orders as follows (listed also in Table 3):
5 x 9.823 = 49.11;  6 x 8.127 = 48.76;  7 x 7.022 = 49.15;  9 x 5.445 = 49.00
Consequently, a complete scan was made to find clusters of orders that matched target "D" spacing values from 30 Angstroms to 140 Angstroms, rising in increments of 0.05Angstroms. Results were accumulated whenever a multiple of one of the 18 first order values matched one of these trial "D" values within less than 0.5% i.e. $D = nd_K$, with $n = 1,2,3...$ where the measured first order spacing is $d_K = \lambda/(2\sin\Theta_K)$.

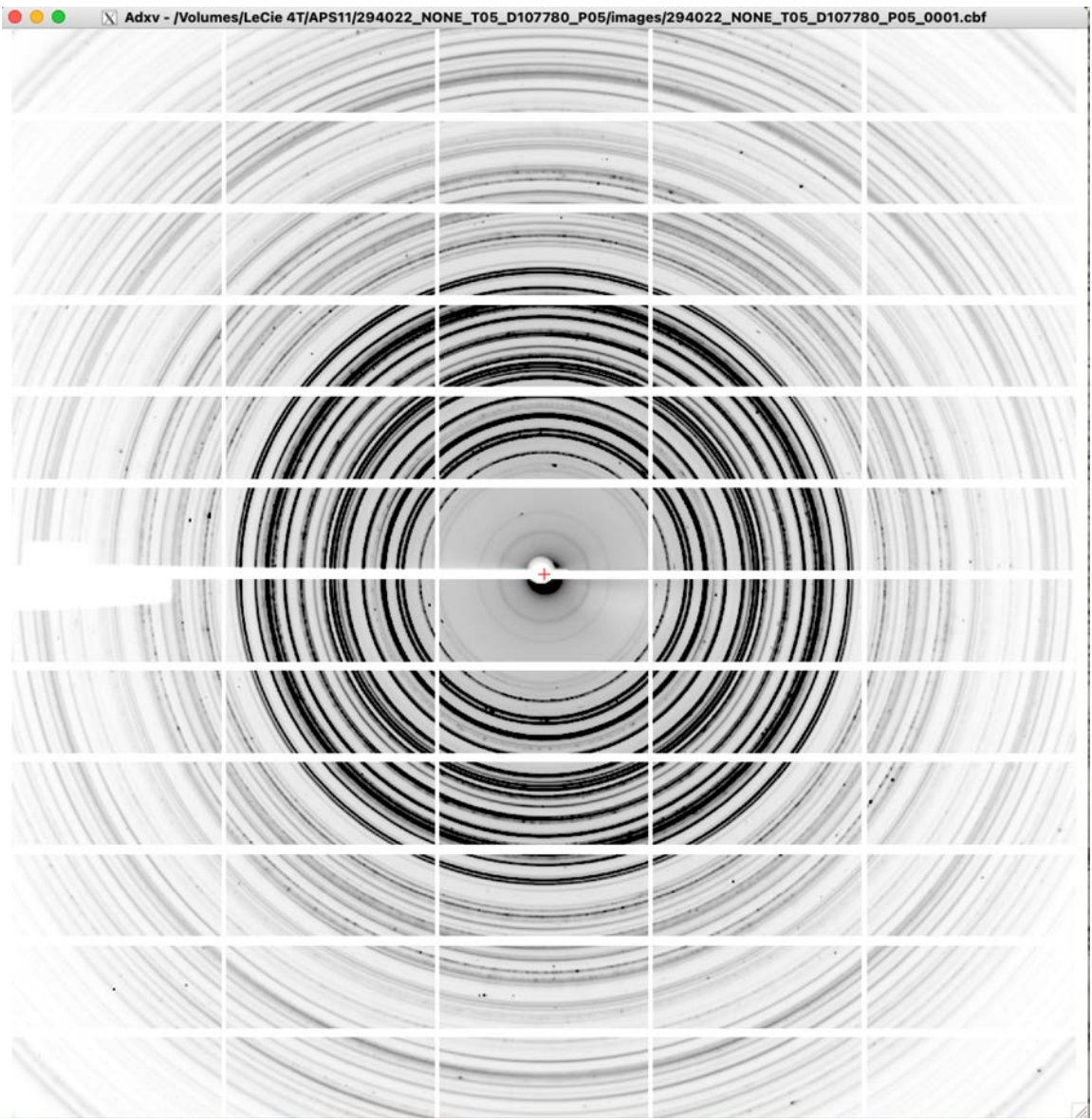

**Figure 7. X-ray diffraction at 2.066 Angstrom from ooid in present era stromatolite showing dark lattice rings between 11.54 and 4.81 Angstroms from the center outward, plus calcium carbonate rings in a faint outer pattern.**

The results of this whole scan are plotted in Figure S2.3 with the most prominent clusters listed here in Table 4. Long sequences of higher order matches prompted the next stage of analysis in which the higher order fits were compared to "D" spacing expectations for the putative diamond 2H lattice [9] of hemoglycin, the detailed calculation of these spacings being laid out in S2. Initially, with an undistorted perfectly tetrahedral lattice there was moderately good agreement, however excellent agreement was obtained with a 4.5 degree increase (to 23.97 ± 0.5 deg) in the angle $\alpha$ between the quasi-hexagonal "sides" and the plane perpendicular to the trigonal symmetry axis [9]. It is concluded that the sample ooid from a recent stromatolite contains an axially distorted diamond 2H open 3D lattice with an inter-vertex spacing of 49.0 ±0.2 Angstroms, and that iron atoms at the vertices provide the strong x ray scattering necessary to observe the lattice. The lattice is filled with calcium carbonate in the crystal forms aragonite and calcite. Because we see rings rather than spots, the ooid comprises a polymer lattice with multiple small crystals in many orientations. The hemoglycin lattice itself is in many orientations, possibly aligned in regions associated with micro-crystals.

**Additional first order diffraction data from crystals and ooids.**

In Section S3 additional diffraction data is given on a) crystals in fossil stromatolite extract and b) ooids from both present day and fossil stromatolite. Out of the fossil samples, only fossil No.2 from Wyoming (provenance in Methods) yielded any ooids. X-ray diffraction results for ooids in the two present day stromatolites and in a second Wyoming sample, No. 2, are compared in Table S3.1. No ooids could be found in Fossil stromatolite No. 1.

In regard to a): In Table S3.1 there was a degree of commonality between different crystal samples of Wyoming Fossil No. 1, and a match between fossil stromatolite and the prior report of diffraction from a fiber crystal of meteorite Acfer 086. The agreement related to the proposed separation of iron atoms at the junction of hemoglycin rods in a rectangular lattice [9].

In regard to b): Table S3.2 compares ring patterns in ooids (provenance in Section S3.2). The Shark Bay ooids are 1. as found, and 2. acid treated as in Methods. The San Salvador ooids were the most recent. Many of the rings were identified as aragonite. Interestingly the 2.1Gya fossil also contained aragonite, implying a "mild" thermal history in view of the tendency for aragonite to transition into more stable calcite at high temperatures [17].

## Section 4. Infrared absorbances of stromatolite ooid and meteoritic material.

**Amide I components as an indicator of secondary structure**

Hemoglycin, primarily composed of antiparallel strands of glycine, has the same peptide backbone as the synthetic polypeptide poly-L-lysine which has been extensively studied. Susi et al. [18] reported infrared absorption measurements of the transition from random conformation through alpha-helix to antiparallel-chain pleated beta sheet. For poly-L-lysine in the random configuration there was a single amide I band centered at 1647cm$^{-1}$ (6.07μm) that was relatively broad at 40cm$^{-1}$. After transition to an antiparallel beta sheet the amide I band was split into two components, the *a-* at 1616cm$^{-1}$ (6.19μm) and the *a+* at 1690cm$^{-1}$ (5.92μm) in the case of H$_2$O solution, each having a relatively narrow width of <20cm$^{-1}$. In that work the related amide II band, typically seen at about 1550cm$^{-1}$ (6.45μm), was suppressed by deuteration of the amide nitrogen [18,19], a consequence of D$_2$O being preferentially used [18] in order to avoid H$_2$O absorption around 6 microns.

The degree of amide I splitting increases with the areal extent of the antiparallel beta sheet, details reported by Cheatum et al. [20]. In ([20], Figure 4) it is found that the splitting is mostly accomplished at the scale of 4 x 4 unit cells in a model, where each unit cell covers four amino acid residues, two in one chain hydrogen-bonded to a pair in an adjacent antiparallel chain. The lower frequency peak *a-*, at 1635cm$^{-1}$ (6.12 μm) in [20], is always much more intense than the up-shifted *a+* peak at 1680cm$^{-1}$ (5.95 μm) in [20]. This splitting does not occur at all in a two-strand antiparallel beta sheet, but always occurs in extended antiparallel beta sheets to the extent that numerous authors have identified the amide I splitting as a reliable marker for the antiparallel beta sheet secondary structure [20,21,22,23]. The calculated strength ratio [23] of *a+* to *a-* is low, as seen experimentally in poly-L-lysine [24] and again in the present data for hemoglycin.

**Analysis of the infrared absorption**

1. Amide bands dominate the mid-IR absorption in both meteorite and ooid samples (Figure 8).

2. Both samples show Amide I splitting into two components: a strong lower frequency component and a very much weaker higher frequency component (Figure 8). These are identified with the extensively researched *a-* and *a+* components, respectively. The degree of splitting depends upon the extent of the sheet and whether it has multiple layers. The SM2 crystal has less extensive sheets, correlated with less splitting [20]. In support of this is the presence in SM2, but not in ooid, of a very strong 3300cm$^{-1}$ (3.03 μm) band due to N-H stretch, implying a large component of N-H groups that are freely vibrating as opposed to being involved in the C=O--H-N hydrogen bond within an extended sheet (Figure S1.2 for illustration of N-H stretch and antiparallel beta sheet).

In view of the above we conclude that each of the substances under test is a polymer of amino acids arranged in an <u>extended</u> antiparallel beta sheet.

3. The amide II band is also seen in both samples, at 1534 cm$^{-1}$ (6.52 μm, ooid) and 1550cm$^{-1}$ (6.45 μm, SM2 crystal). The Amide II band of hemoglycin in ooid is more intense than amide I, contrary to the observation in SM2 and to the approximately 0.5 ratio expected for amide II/amide I in an extended antiparallel beta sheet [23,25]. The reason for this anomalous Amide II intensity in ooid is not known. The preparation of the ooid sample involved dilute acetic acid treatment to free the polymer from a matrix of calcium carbonate.

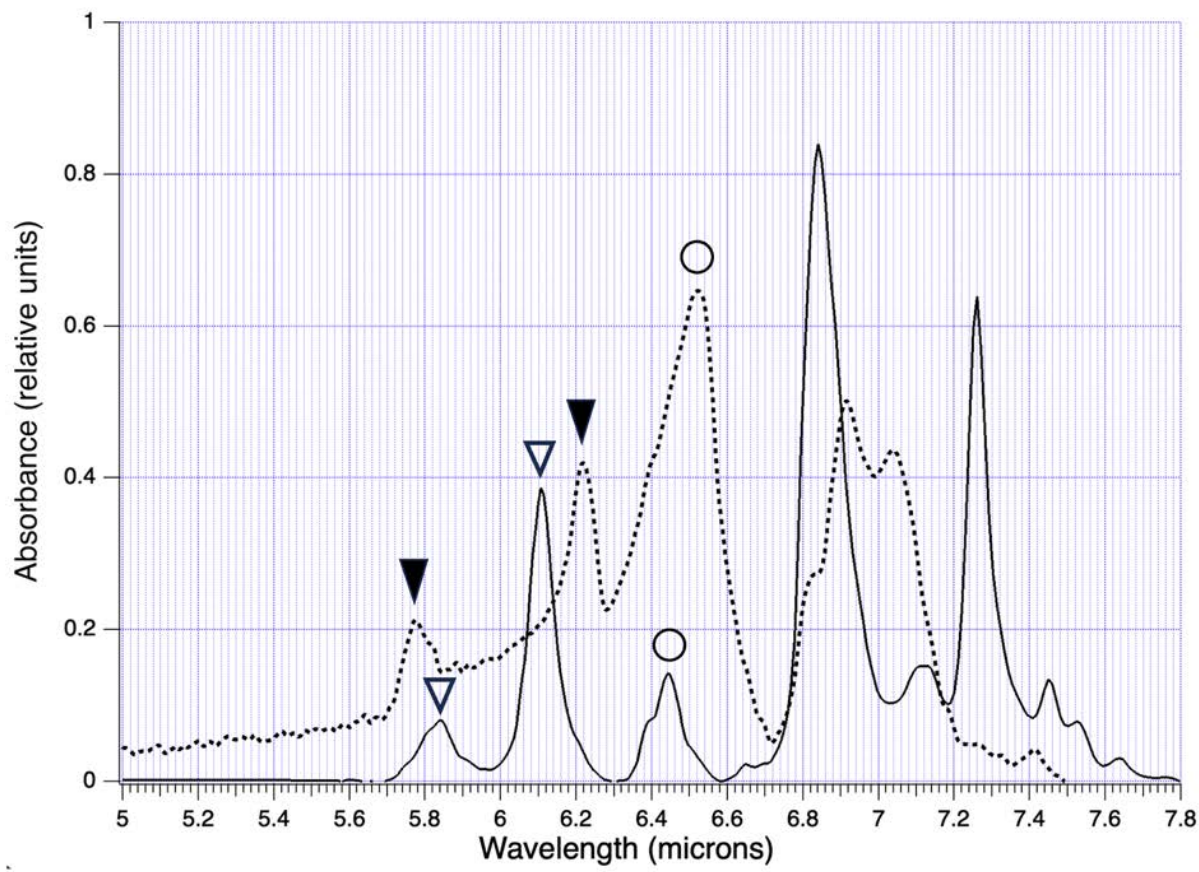

**Figure 8. Mid-infrared absorption spectra of Meteorite crystal SM2 (full line) and Shark Bay stromatolite ooid (dotted curve). Amide I components open triangles (SM2) and filled triangles (ooid). Amide II components circles. Additional lines from 6.8microns to 7.3microns discussed in text.**

In the ooid data of Figure 8 the complex of bands between 6.8 μm and 7.1 μm (dotted curve) is due to amorphous calcium carbonate [26], additionally identified by its diminution with longer acid treatment. In the same region the meteoritic spectrum has strong and relatively sharp absorptions at 1462cm$^{-1}$ (6.84 μm) and 1377cm$^{-1}$ (7.26 μm) that are not spatially correlated with the amide band absorptions within the crystal. They are from a non-amide source that has been identified [27] as most probably saturated hydrocarbons, the closest match being to pentane, which has $CH_2$ deformation bands at 1470cm$^{-1}$ (6.80μm) and 1378cm$^{-1}$ (7.26 μm). Saturated hydrocarbons may constitute more than 30% of all the carbon in the line of sight to these protostar sources [27]. We emphasize that hemoglycin in material ultimately derived from a molecular cloud has hydrocarbon bands that are known to be dominant in the interstellar medium, whereas hemoglycin in carbonate minerals within the ocean has no hint of saturated hydrocarbons.

## Discussion

We find that four very different lines of evidence all point to the presence in stromatolites of the hemoglycin polymer previously only known from meteorites. In the first 2.1Gya fossil stromatolite sample the fraction of heavy isotopes in the core unit mass spectrometry peak complex is comparable to that in the Orgueil meteorite, which seems to indicate that there was preservation of in-fall hemoglycin in the predominantly calcium carbonate fossil. Following studies of the Barberton Greenstone Belt of South Africa, Lowe et al. [28] have discussed revision of the temporal profile of meteoritic in-fall toward an ongoing, more gradual decline [29] than in the Late Heavy Bombardment theory in which in-fall declined abruptly to present day levels at about 3.9Gya. The in-fall rate at 2.5Gya, the end of the Archean, may have been 10 times greater than at present. Most of the in-falling material would likely resemble the Orgueil meteorite, which is known to be characteristic of solar system material [30]. Our present work with Orgueil has shown that relatively complex chemicals such as hemoglycin can survive in-fall, although not, presumably, in the heat and shock conditions of a major impact.

Calcium carbonate ooids are the primary mineral constituent of present-day stromatolites. However, through geological time there can be partial replacement of calcium carbonate in ooids by silicates, as seen by for example [31] in 2.72Gya fossil ooids from Western Australia.

The survival of hemoglycin is attributed to its being an incredibly tough molecule, that, once formed in a protoplanetary disc, often becomes internally mineralized, remaining within the mineral as an extensive low density lattice. This state has now been observed for the first time in x ray analysis at 2Angstroms. In the protoplanetary disc, molecules circulate from hot, high ultraviolet, to cold regions and should an open hemoglycin lattice get near the new sun it would degrade and never feature as an in-fall polymer. Hemoglycin is found in many meteorites indicating that some hemoglycin from the colder regions of the disc does persist to seed planets via in-fall. Planets forming from a protoplanetary disc via gravity, as they enlarge, experience heating due to Al isotope decay, meaning that hemoglycin landing on such an early entity will not remain intact. A planet like Earth conducive to developing complex chemistry, going on to life forms, will rely on asteroid in-fall once it is sufficiently cool for the molecules to survive.

Beginning approximately 2.4Gya and reaching partial completion 2Gya, there was a build-up of oxygen in Earth's atmosphere from almost zero to a fraction of the present day value, known as the great oxygenation event (GOE) [32]. It has been proposed that gradually cyanobacteria dominated over anoxygenic photosynthetic bacteria, in a process dependent on geochemical changes together with locally increasing sources of oxygen [32]. However ultraviolet radiation reaching the Earth's surface prior to this event was as much as 400 times greater than in the present day [33,34]. This was in the UV band approximately between 200nm and 300nm which is the most destructive to nucleotides, equivalent to an E. Coli mutation doubling dose every quarter second [33], much too high for any organism to survive. It requires several metres of water to attenuate the pre-GOE radiation levels to equate with present day surface exposure. The question is whether the earliest organisms were able to constitute themselves and thrive sufficiently to perform global oxidation in spite of an extremely hostile surface environment. Out of the present work comes an alternate hypothesis related to the possibility of reliable abiotic water splitting by hemoglycin in the presence of sunlight.

We have done initial quantum chemical modeling on a water-splitting reaction that hemoglycin can engage in, via direct absorption of UV light. We believe that it is a two-step reaction cycle that goes via
1. hemoglycin + $H_2O$ + h$\nu$ -> hemoglycin (OH) + $H_2$
2. hemoglycin(OH) + $H_2O$ + h$\nu$ -> hemoglycin + $H_2O_2$

followed by the release of $O_2$ from $H_2O_2$.
There are no other participants in this process, no catalysts and "room temperature" operation. The finding of hemoglycin in a fossil stromatolite therefore opens up the possibility that its (proposed) oxygen producing ability could have "kick-started" the GOE, producing an increasing degree of ultraviolet protection for complex biology. Furthermore, it could potentially provide chemical energy to its surroundings.

The R-chirality of hemoglycin that lets it absorb visible light [11] sets it apart from S-based life involving amino acid protein. This may be its most important property allowing separation of systems with hemoglycin that are essentially abiotic to be very distinct from biochemical systems. On early Earth this system divide could have been maintained with fossil stromatolites forming their mineral parts abiotically [2,35], in possible contrast to present day stromatolites [1]. However, there is evidence of organic material comparable to that in present day stromatolites having been present in the neo-Archean [31]. It would be of great interest to know whether hemoglycin present in modern day ooids still provides an energy source.

A 2023 report emphasizes that UV-driven chemistry in protoplanetary disks is a "signpost" for planet formation [36]. The hemoglycin polymer is not referenced in [36] with its response at both 480nm and 6μm [11] but at least the factor of light on molecules is raised. In 2022 we suggested that hemoglycin was a factor in accretion [11]. To acknowledge the need for an abiotic factor, in this case light, acting on a molecule that allows energy transfer is a step forward to understanding paths to the evolution of accreted matter destined for planet formation and for biochemical evolution. A report has been published that hemoglcyin is the likely molecule causing absorption in molecular clouds from 3-13.2microns with strong peaks at 6.2microns [37].

## Conclusions

Modern day stromatolite ooids and fossil stromatolite (2.1Gya) from the Medicine Bow Mountains of Wyoming contain hemoglycin, the space polymer. At 2.1Gya there was ongoing substantial asteroidal delivery, including hemoglycin, to Earth where water was present as tidal pools or early oceans, conditions that support stromatolite formation. Light is the important agent for hemoglycin in that it allows the molecule to potentially pass on energy to other chemistry, and possibly lead to atmospheric oxidation. Once the early Earth had hemoglycin it had solar driven chemical energy that may have led by paths unknown and to be investigated, to the first life forms, the stromatolites and their bacterial mats. The first events together (hemoglycin in-fall and formation of first stromatolites) could have been abiotic and may have preceded simple organisms like cyanobacteria.

**ACKNOWLEDGMENTS**

We wish to thank the late Guido Guidotti of Harvard who gave encouragement and advice for this extra-terrestrial polymer research. We thank Charles H. Langmuir and Zhongxing Chen of the Department of Earth and Planetary Science, Harvard, for use of their Hoffman clean room facilities, and Sunia Trauger, the senior director of the Harvard center for MALDI mass spectrometry. The stromatolite FTIR measurements were performed at Harvard Center for Nanoscale systems (CNS). This research used two synchrotron resources: 1) The Advanced Photon Source, a U.S. Department of Energy (DOE) Office of Science User Facility operated for the DOE Office of Science by Argonne National Laboratory under Contract No. DE-AC02-06CH11357. Use of the Lilly Research Laboratories Collaborative Access Team (LRL-CAT) beamline at Sector 31 of the Advanced Photon Source was provided by Eli Lilly and Company, which operates the facility. 2) The Diamond Light Source, beamline I24 and B22 (for meteorite FTIR), Harwell Science and Innovation Campus, Didcot, OX11 0DE, UK. The Orgueil meteorite samples n234, dispatch number ar813, Colhelper request number 170600, were provided by the Museum National D'Histoire Naturelle (MNHN) Paris by Beatrice Doisneau. We particularly thank BD for sending Orgueil samples that had high IOM. We thank Cfa, Harvard and Smithsonian for supporting the mass spectrometry analysis. Andrew Knoll of OEB, Harvard and Museum of Comparative Zoology, 26 Oxford Street, Cambridge, MA 02138 provided the present-day stromatolite samples from San Salvador and from Hamlin Pool, Shark Bay Australia, collected by Elso Barghoorn in 1971.

**DATA AVAILABILITY**

The data that support the findings of this study are available from the corresponding author upon reasonable request and at the Harvard Dataverse repository via URL: https://doi.org/10.7910/DVN/HRU9SJ, once accepted.

31. Flannery, D. T., Allwood, A. C., Hodyss, R., et al., "Microbially influenced formation of Neoarchean ooids", *Geobiology*, **1-10** (2018).

32. Olejarz, J., Iwasa, Y., Knoll, A. H. and Nowak, M. A. "The Great Oxygenation Event as a consequence of ecological dynamics modulated by planetary change", *Nature Communications* **12**, 3985 (2021).

33. Cockell, C. S. "The ultraviolet history of the terrestrial planets – implications for biological evolution", *Planetary and Space Science* **48**, 203-214 (2000).

34. Karam, P. A. "Inconstant Sun: How Solar evolution has affected cosmic and ultraviolet radiation exposure over the history of life on Earth." *Health Physics* **84**, 322-333 (2003).

35. Grotzinger, J. P. and Rothman, D. H. "An abiotic model for stromatolite morphogenesis", *Nature* **383** 424-435 (1996). (Cowles Lake Formation, Wopmay orogen, northwest Canada, age 1.9Gya).

36. Calahan, J. K., Bergin, E. A., Bosman, A. D. et al. "UV-driven chemistry as a signpost of late-stage planet formation". *Nature Astron.* **7** 49-56 (2023). https://doi.org/10.1038/s41550-022-01831-8

37. McGeoch, J. E. M and McGeoch, M.W. (2024) Polymer amide as a source of the cosmic 6.2 micron emission and absorption arXiv:2309.14914 [astro-ph.GA]. MNRAS DOI: 10.1093/mnras/stae756.

# Supplementary information to Fossil and present-day stromatolite ooids contain a meteoritic polymer of glycine and iron.

[1*]Julie E M McGeoch, [2]Anton J Frommelt, [3]Robin L Owen, [4]Gianfelice Cinque, [5]Arthur McClelland, [6]David Lageson and [7]Malcolm W McGeoch

[1]Department of Molecular and Cellular Biology, Harvard University, 52 Oxford St., Cambridge MA 02138, USA & High Energy Physics Div, Smithsonian Astrophysical Observatory Center for Astrophysics Harvard & Smithsonian, 60 Garden St, Cambridge MA 02138, USA.
[2]LRL-CAT, Eli Lilly and Company, Advanced Photon Source, Argonne National Laboratory, 9700 S. Cass Avenue, Lemont, IL, 60439
[3,4]Diamond Light Source, Harwell Science and Innovation Campus, Didcot, OX11 0DE, UK.
[5]Center for Nanoscale Systems, Harvard University, 11 Oxford St, LISE G40, Cambridge, MA 02138, USA.
[6]Department of Earth Sciences, 226 Traphagen Hall, P.O. Box 173480 Montana State University, Bozeman, MT 59717.
[7]PLEX Corporation, 275 Martine St., Suite 100, Fall River, MA 02723, USA.
*Corresponding author. E-mail: Julie.mcgeoch@cfa.harvard.edu

## Section S1: Current Status of Structural Knowledge on Hemoglycin

The structure of hemoglycin has revealed itself in a diverse set of measurements spanning from 2015 up to the present day. Some aspects, listed below, need further confirmation. To date, hemoglycin has only been available in micro-gram quantities, and even then has not been of guaranteed purity. It crystallizes in many forms that appear to be determined by the presence and nature of additional substances that have accreted within the several types of open lattice that it can form. We summarize here the principal findings on structure. Figure S1.1 depicts the four structural ingredients which combine in different sets to make all crystal and lattice types. There is a "core unit" of mass 1494Da that comprises two antiparallel 11-residue polyglycine strands, with hydroxylation of two residues on each strand, closed out by iron atoms. Prominent in the mass spectrum of meteoritic extracts [1] is a 1638Da variant in which the core unit has two additional Fe atoms, plus two additional oxygen atoms. Beyond these are silicon connector atoms that exhibit two-fold or fourfold valence bonding to oxygen at the ends of the 1494 and 1638Da "rods". One structural example is the tetrahedral junction shown in Figure S1.1 that is repeated throughout a three-dimensional lattice that hemoglycin forms, with vertices related to each other in diamond 2H symmetry ([2], and present MS).

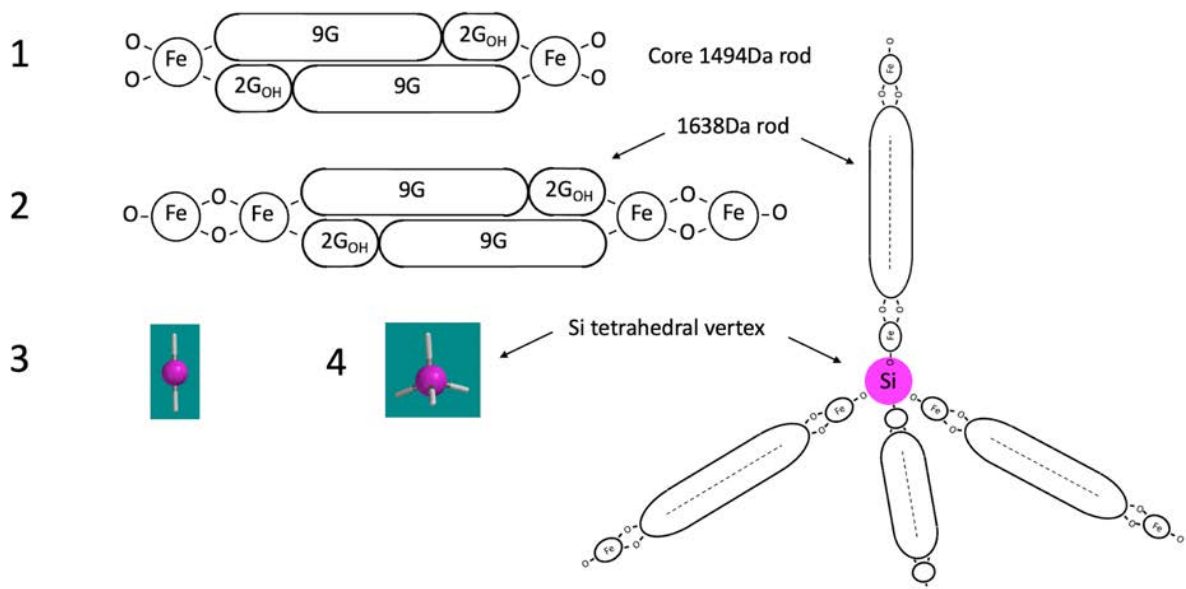

**FIGURE S1.1. The four building blocks of hemoglycin extended structures. 1. a 1494Da core unit; 2. a 1638Da rod containing 4 Fe atoms; 3. Si atoms with two bonds; 4. Si atoms with 4 bonds. As an example, the tetrahedral junction of the 3-D hemoglycin lattice is shown.**

In Figure S1.2, three 1494Da "core units" of hemoglycin are shown hydrogen-bonded edge to edge, another example of an assemblage based on components in Figure S1.1. The entity shown has hydrogen on each terminal oxygen to serve as a place-holder for oxygen-oxygen, or oxygen-silicon bonds in extended two- and three- dimensional lattices. The structure in Figure S1.2 is based upon equilibrium geometry B3LYP energy minimization within each unit, followed by MMFF energy minimization as a trimeric sheet.

Structural findings are now listed:

1. Polymers containing glycine and hydroxyglycine, ranging from 2 to >20 residues, were first observed in MALDI mass spectrometry within an extract of the Allende meteorite [3].

2. A 4641Da molecule was observed in MALDI in both Allende and Acfer 086 extracts [4]. The details of this peak complex were different in each case, and there was zero signal at the same mass in a volcano control processed in the same way. Higher multiples up to 4 units of 4641Da were present, indicating that the species was connecting into larger assemblies. Lower mass fragments contained glycine and hydroxyglycine, with high isotope enhancements that confirmed extraterrestrial origin. This was the first observation of an organized entity as opposed to random-length polymers. Later, once the core unit was established, the 4641 entity was identified as a triskelion of 1494Da core units connected by silicon bridges [1].

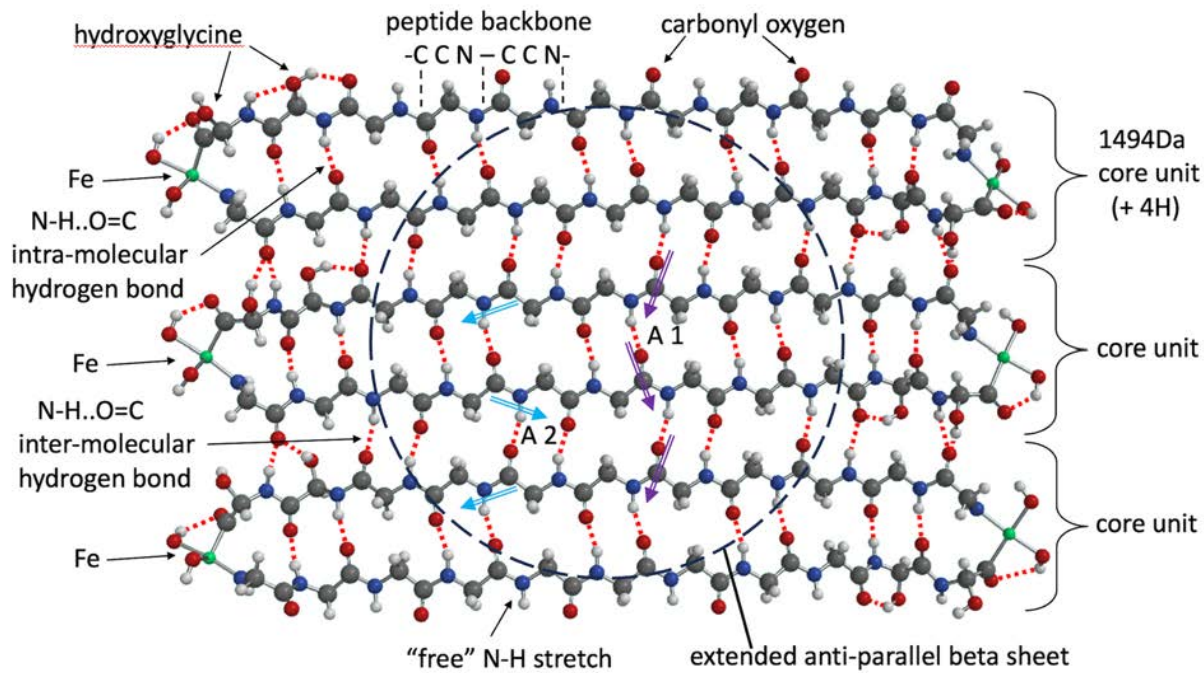

**Figure S1.2. Array of 3 hemoglycin "core units" bonded in a sheet by inter-molecular hydrogen bonds. Iron atoms green, hydrogen white, carbon grey, nitrogen blue and oxygen red. Hydrogen bonds red dotted lines. Energy minimized in MMFF. Amide I dipoles (A1) and Amide II dipoles (A2) shown, on average perpendicular to each other.**

3. In MALDI [1] an extract of Acfer 086 yielded dominant 1566Da and 1638Da peaks (seen at 1567 m/z and 1639 m/z in mass spectra). For the first time iron was detected in the molecule via its $^{54}$Fe content that added subordinate peaks at (-1) and (-2) mass units below the principal "mono-isotopic" peak. The amplitude of these subordinate peaks indicated, for example, that the 1566Da entity contained two Fe atoms, while the 1638Da entity contained four Fe atoms. When the principal peaks were subjected to MS/MS analysis, in which they were broken into fragments, the larger m/z peaks yielded lower mass peaks from the original spectrum, indicating the simultaneous presence of a progressive series that comprised a basic or "core" unit with various additions. Crystals of extracts from the Orgueil meteorite that potentially contained a purer and more uniform molecular species (this MS), yielded the 1494Da core unit peak predicted in [1].

4. A FIB/TOF/SIMS experiment [5] on Allende and Acfer 086 micron-scale particles revealed that $^{15}$N was enriched at a cometary level in components with the same fragmentation pattern as a polymer amide backbone of –CCN-CCN-. This $^{15}$N/$^{14}$N ratio was used, together with total isotope enrichment from mass spectrometry, to determine the D/H ratio of hemoglycin to be 25 times terrestrial [1] and, when applied to data in the present MS, as high as 40 times terrestrial.

5. X-ray diffraction of a fiber crystal of Acfer 086 [2] revealed a set of diffraction orders with a 4.8 nm fundamental spacing. This matched the spacing of Fe atoms in computed hemoglycin lattices with silicon junctions [2] confirming the length of hemoglycin, which had previously only been calculated using quantum chemistry. Later, a crystal from the Sutter's Mill meteorite yielded an even more extended "ladder" of high diffraction orders (up to N=12) and gave an almost identical Fe-Fe spacing. Such high diffraction orders are seen because Fe atoms act as "markers" that have much higher X-ray scattering than the H,C,N,O constituents of the polymer strands. Furthermore, several Fe atoms can be in close proximity at the vertices of extended polymer lattices, enhancing the effect.

6. Following the prediction of a chiral 480nm absorption in hemoglycin, a Sutter's Mill crystal exhibited the predicted absorption [6]. The absorption was associated with the Fe-glycine bond region, being strong at 480nm only when a) there was a glycine C-terminus next to Fe, and b) that glycine residue carried an 'R' chirality hydroxylation on its alpha carbon atom. The existence of the absorption simultaneously confirmed the presence of Fe in the molecule and the location of at least one hydroxylation adjacent to the Fe atom.

7. In the present MS the structure of the three-dimensional space-filling lattice [2] was confirmed in 2-Angstrom X-ray scattering of a present-day ooid. We report 18 higher order diffraction rings that confirm the diamond 2H structure [2] with its tetrahedral lattice junction (Figure S1.1) and again show an inter-vertex spacing of 4.9 nm, consistent with the calculated length of the 1638Da rods, including Si atoms that join the rods.

8. Infrared (IR) absorption, first presented in the present MS, confirms that hemoglycin crystals contain anti-parallel beta sheets, as first suggested in [1,4]. In both ooid and meteorite the Amide I band was split indicating an <u>extended</u> anti-parallel beta sheet, as shown in Figure S1.2 (absorbance data in the main text Fig.8). This splitting would not be expected in solitary core units. For the first time, via Fourier transform IR spectrometry, the secondary structure of this space polymer has been seen and is found to be in accord with the above identification of an amino acid polymer of glycine.

9. Observation of a visible fluorescence induced by X-rays [7] confirmed that the 480nm absorption had a counterpart in molecular fluorescence. The fluorescence of ooids in the X-ray beam comprised the same set of components as for an Orgueil meteorite crystal, although their amplitudes were different, reflecting a different chemical history (present MS).

Further work is needed to:
a) determine the location of hydroxyl groups beyond the proven set on residues with C terminus adjacent to Fe atoms. They could be on the second residue distant from Fe, in view of possible photon-induced reactions associated with the visible absorption, or on the N terminus residue adjacent to Fe.
b) study silicon, which is believed to be present at most types of lattice junction [1,2], bonded to oxygen atoms on the terminal Fe and imparting tetrahedral symmetry to the junctions in the diamond 2H three-dimensional lattice. (Figure S1.1). However, the mass of Si has only appeared as a necessary part of the mass fit to the 4641 Da triskelion [1]. It is desirable to confirm the presence of silicon more generally at the junctions of 2D and 3D lattices because it is apparently necessary, in addition to rod lengths, to provide the 4.9nm inter-vertex spacing that is repeatedly observed in diffraction from 2D and 3D lattices. The core unit itself as shown in Figure S1.1 is 4.0nm in length between iron atoms.

**References for all S sections**

1. McGeoch, M. W., Dikler, S. and McGeoch, J. E. M. "Meteoritic proteins with glycine, iron and lithium", arXiv:2102.10700 (2021).
2. McGeoch, J. E. M. and McGeoch, M. W., "Structural organization of space polymers", *Phys. Fluids* **33**, 067118 (2021).
3. McGeoch, J. E. M. and McGeoch, M. W. "Polymer amide in the Allende and Murchison meteorites". *Meteorit and Planet Sci.* **50**, 1971-1983 (2015).
4. McGeoch J. E. M. and McGeoch M. W. (2017) . A 4641Da polymer of amino acids in Acfer-086 and Allende meteorites. https://arxiv.org/pdf/1707.09080.pdf
5. McGeoch, M. W., Samoril T., Zapotok D. and McGeoch J. E. M. (2023) Polymer amide as a carrier of $^{15}N$ in Allende and Acfer 086 meteorites. Under review at *International Journal of Astrobiology*.
6. McGeoch, J. E. M. and McGeoch, M. W., "Chiral 480nm absorption in the hemoglycin space polymer: a possible link to replication", *Scientific Reports*, **12**:16198 (2022), https://doi.org/10.1038/s41598-022-21043-4
7. McGeoch, M. W., Owen, R. L., Jaho, S. and McGeoch, J. E. M. "Hemoglycin visible fluorescence induced by x rays", *J. Chem. Phys.* **158**, 114901 (2023).
8. XQuartz 2.8.5, X window system, 2003-2023 X.org Foundation, Inc.
9. McGeoch, M. W. and McGeoch, J. E. M. "Hexagonal Cladding of a Hemoglycin Vesicle in X-ray Diffraction", Unpublished

## Section S2: Lattice analysis via higher order diffraction

In X-ray analysis of meteorite polymers of amino acids we observe high order diffraction "ladders" in thin samples that are sheet-like [6], or fibers apparently composed of rolled up sheets [2]. In addition, there is evidence [ref.2 +unpublished lattice diffraction] for a truly three-dimensional lattice of the diamond 2H structure that represents the maximum volume that a 3D lattice of identical "rods" can enclose [2]. The rods in this case are polymers of glycine [1,2] comprising anti-parallel glycine chains of 11-residue length closed at each end by an iron atom [1] with hydroxylation of glycine residues adjacent to Fe, termed hemoglycin. X-ray diffraction from such lattices is from two main features: a) Fe atoms in groups at vertices, the groups spaced from each other by the 49A length of a rod [2, 6] and b) nano-crystals of any substance filling the lattice spaces.

Stromatolites are dominantly formed of ooids, which are small ovoid, predominantly calcium carbonate, grains of typical length a few hundred microns. Following the mass spectrometry finding that the 1494Da core unit of hemoglycin was present in both the Orgueil meteorite and fossil stromatolite, an X-ray study was made of modern day ooids found within a recent stromatolite sample from Shark Bay, Australia. Two X-ray wavelengths, 0.979 Angstroms and 2.066 Angstroms, were used on APS beam line 31-1D-D. These lay respectively at higher energy and lower energy than the Fe K edge absorption at 7.1keV (1.74 Angstroms). Differences in X-ray scattering between these two wavelengths would potentially give information on the disposition of Fe atoms within the ooid. The diffraction patterns were indeed markedly different. At the shorter wavelength, a set of rings between 3.86 and 1.61 Angstroms, listed in Table S2.1, indicated that both the calcite and aragonite forms of calcium carbonate were present. No significant rings were observed at "d" spacing larger than 3.85 Angstroms.

**Table S2.1. Ooid (Shark Bay) diffraction rings on APS at 0.979 Angstroms indicating the presence of calcite and aragonite, labeled w, m, s, vs for weak, medium, strong, very strong. The run was at nominal room temperature.**

| Strength | "d" spacing | Aragonite (a) Calcite (c) |
|---|---|---|
| w | 3.85 | c |
| s | 3.41(5) | a |
| s | 3.29 | a |
| vs | 3.02 | c |
| s | 2.71(5) | a |
| s | 2.49 | c, a |
| w | 2.42 | - |
| m | 2.38 | a |
| s | 2.34(5) | a |
| m | 2.28(5) | c |
| w | 2.20 | - |
| w | 2.11(5) | - |
| w | 2.09(5) | c |
| s | 1.98(5) | a |
| s | 1.91(5) | c |
| vs | 1.88(5) | a |
| s | 1.82(5) | - |
| s | 1.74(5) | a |
| s | 1.72(5) | - |
| w | 1.61 | c |

In contrast, at 2.066 Angstroms, (Figure S2.1) an intense new set of large "d" spacing rings (central) was superimposed upon a weak version of the calcite and aragonite rings (outside), the latter having "d" spacing less than 3.86 Angstroms. The new set of 18 large "*d*" spacing rings ranged from 4.808 to 11.54 Angstroms, <u>when interpreted as first order</u> (Figure S2.1 and data Table S2.2). We show here that these come from higher order diffraction off Fe groupings at the 5nm - spaced junctions of the diamond 2H open lattice structure already identified in meteoritic work [2,6]. The present analysis shows that there is a slight distortion to the lattice involving a relative increase of the trigonal axis.

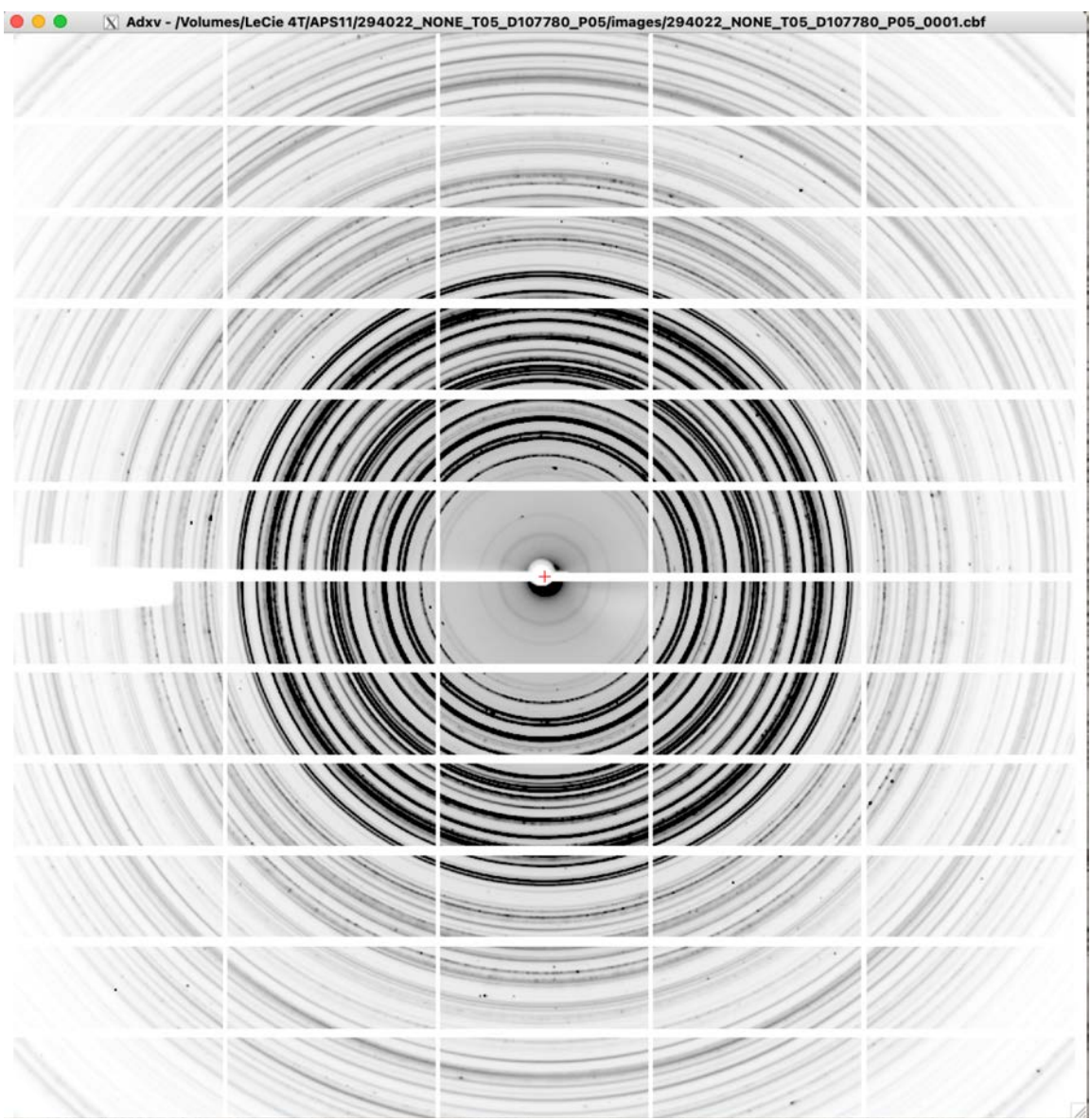

**Figure S2.1 X-ray diffraction at 2.066 Angstrom from ooid in present era stromatolite (Shark Bay) showing dark lattice rings between 11.54 and 4.81 Angstroms from the center outward, plus calcium carbonate rings in a faint outer pattern.**

**Data reduction to lattice "d" spacings from the set of high order rings**
The inner rings in Figure S2.1 had the "d" parameters listed in Table S2.2. These were calculated automatically in XQuartz [4] on the assumption that they were in first order diffraction according to $2d\sin\Theta = n\lambda$ where $n$ is the diffraction order, $\Theta$ is one half of the deflection angle and $\lambda$ is the wavelength. In this data the wavelength was 2.066 Angstroms. The left hand column of Table S2.2 contains the set of 18 rings that represented larger "d" spacing than 4 Angstroms. Their relative intensities are shown in the scan of Figure S2.2, part A. These rings did not match either the calcite or aragonite values, but were reminiscent of the ladders of high order diffraction previously seen in hemoglycin lattices [2, 6]. In [2] the ladder contained orders 2 through 5, with a fitted first order lattice parameter of 48.38 ± 0.2 Angstroms. In [6] the ladder contained orders 2 through 12, with a fitted first order parameter of 49.03 ± 0.18 Angstroms. Quick inspection of the present 18 ring set yielded a fit to 49.0 Angstroms in 5th 6th 7th and 9th orders as follows:
5 x 9.823 Angstroms = 49.11;
6 x 8.127 Angstroms = 48.76 ;
7 x 7.022 Angstroms = 49.15
9 x 5.445 Angstroms = 49.00

When several additional sequences also fitted calculated inter-vertex distances in the ($h$ = 49 Angstrom, diamond 2H) 3D lattice, a thorough programmed search was performed in which each of the 18 data values was assessed as a divisor of trial "$d$" spacings in the range 30 Angstroms to 140 Angstroms, rising in 0.05 Angstrom increments. The number of such divisors found in each 0.05 Angstrom region was plotted in the range 30 to 140 Angstroms in Figure S2.3, after binomial smoothing and raising to the second power to accentuate the longer sequences.

This discovery routine yielded about 10 principal candidates for "$d$" spacings, the main ones being listed in Table S2.1 together with the relevant diffraction orders and the accuracy of a match (always better than 0.5%). Initially not all of these strong candidates matched calculated vertex-to-vertex spacings [2] for the ideal diamond 2H lattice. A mathematical model was then constructed to handle a non-ideal lattice with a constant rod length "h" and variable deviations from the tetrahedral angle, that is, changes to the angle $\alpha$ where (90 + $\alpha$ = 109.471221 deg. is the tetrahedral angle). Changes to $\alpha$ represent axial stretch or compression of the lattice. To anticipate the results, we are able to fit the principal "$d$" spacing candidates with an increase of $\alpha$ by 4.5 deg., computed results shown by red bars in Figure S2.3.

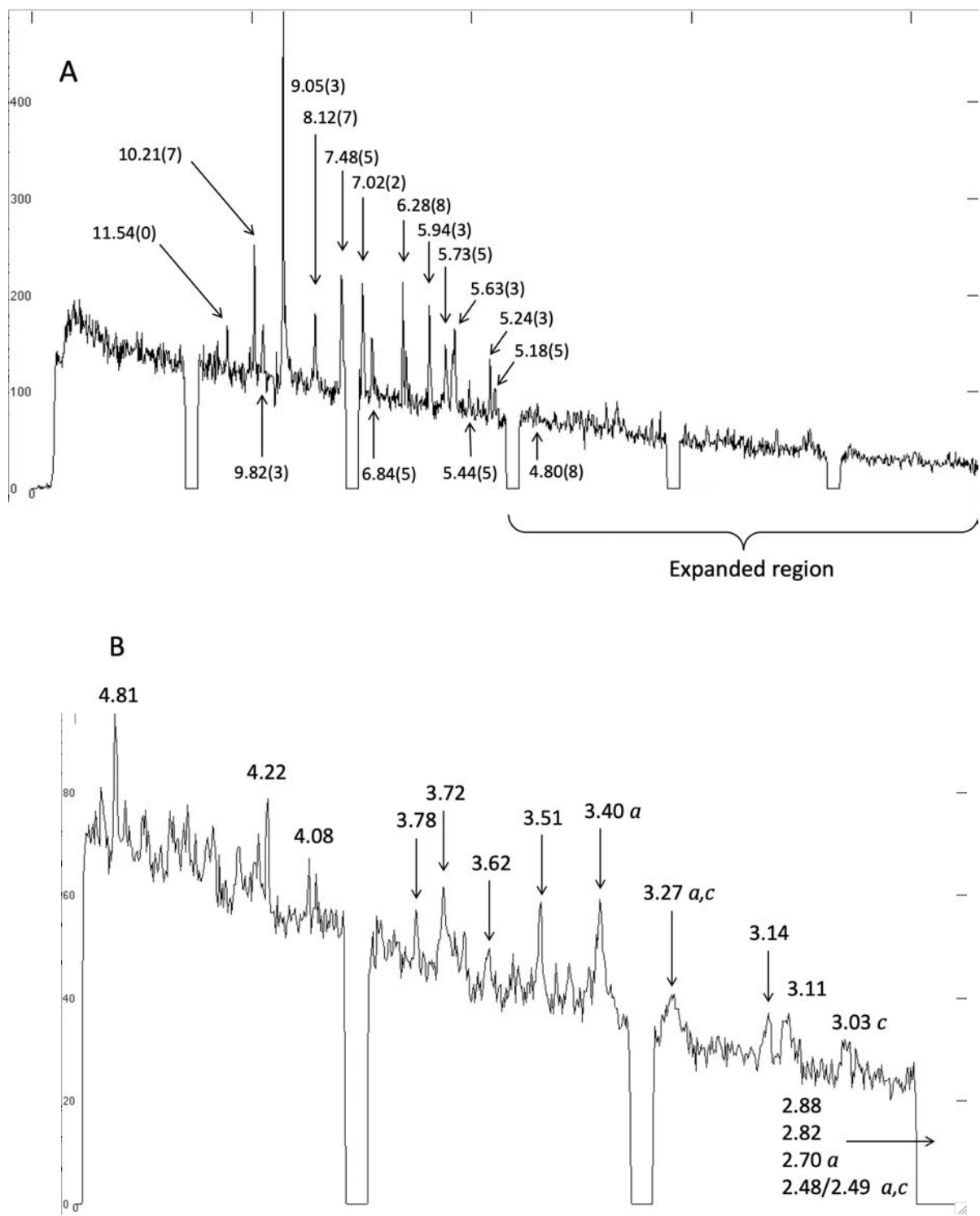

**FIGURE S2.2** A: Vertical intensity scan of Figure S2.1. Note that the 7.12(8) peak is obscured by a detector grid. B: Scan of expanded region. Additional peaks 2.88-2.48 Angstroms were measured on the diagonal.

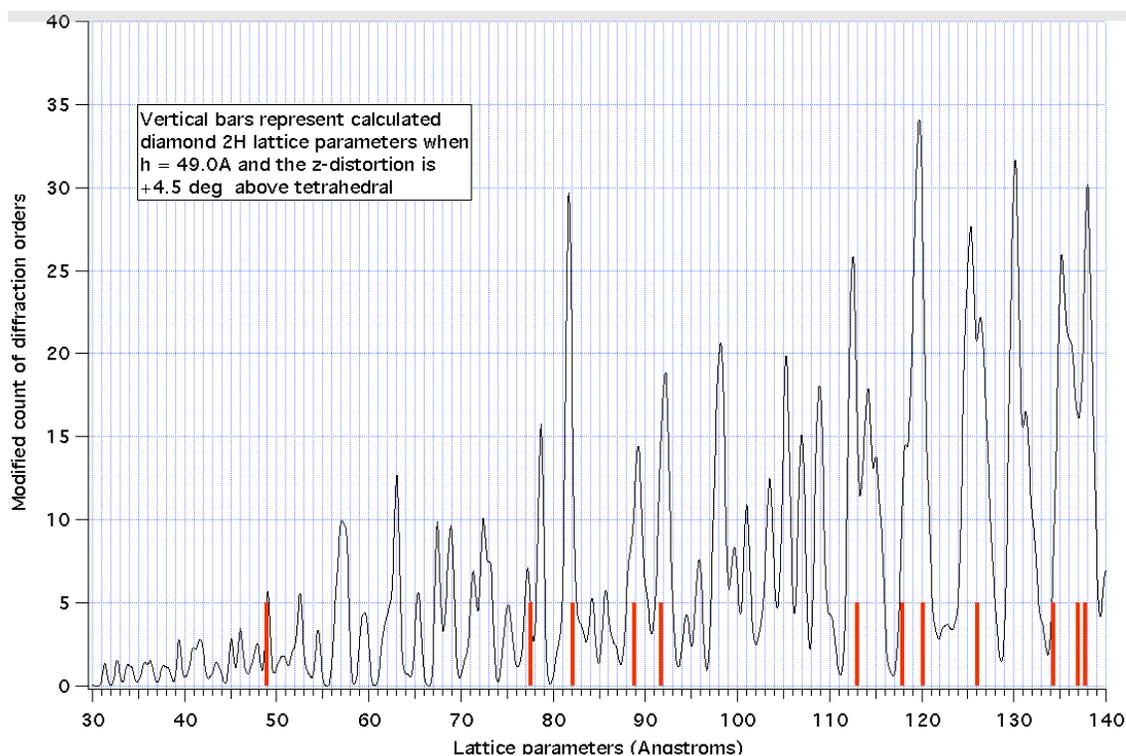

**Figure S2.3.** (thin black line) The results of a computer search for higher order clusters or loci, represented as the square of the number of "hits" within any 0.05 Angstrom band, with binomial smoothing applied. The red bars are calculated "d" spacings for a slightly distorted diamond 2H lattice, discussed below.

**Table S2.2** Ooid diffraction rings in first order (left hand column). Higher order fits (top row) listed as diffraction order in bold with percentage mis-match.

| Angstroms | **49.0** | **81.65** | **92.05** | **112.75** | **119.9** | **126.25** |
|---|---|---|---|---|---|---|
| 4.80(8) |  | **17**, 0.1% |  |  | **25**, 0.2% |  |
| 5.18(5) |  |  |  |  |  |  |
| 5.24(3) |  |  | **17**, 0.1% |  |  | **24**, 0.3% |
| 5.44(5) | **9,** 0% | **15**, 0.2% |  |  | **22**, 0.1% |  |
| 5.63(3) |  |  |  | **20**, 0.1% |  |  |
| 5.73(5) |  |  |  |  | **21**, 0.4% | **22**, 0.1% |
| 5.94(3) |  |  | **15**, 0.1% | **19**, 0.1% |  |  |
| 6.28(8) |  | **13**, 0.4% |  | **18**, 0.4% | **19**, 0.3% | **20**, 0.4% |
| 6.57(8) |  |  |  |  |  |  |
| 6.84(5) |  |  | **13**, 0.1% |  |  |  |
| 7.02(2) | **7,** 0.3% |  |  | **16**, 0.3% | **17**, 0.4% | **18**, 0.1% |
| 7.12(8) |  |  |  |  |  |  |
| 7.48(5) |  |  |  | **15**, 0.4% | **16**, 0.1% |  |
| 8.12(7) | **6,** 0.5% | **10**, 0.1% | **11**, 0.3% |  |  |  |
| 9.05(3) |  | **9**, 0.2% |  |  |  | **14**, 0.3% |
| 9.82(3) | **5,** 0.2% |  |  |  |  |  |
| 10.21(7) |  | **8**, 0.1% |  | **11**, 0.3% |  |  |
| 11.54(0) |  |  |  |  |  |  |

**Mathematical construction of the diamond 2H lattice**

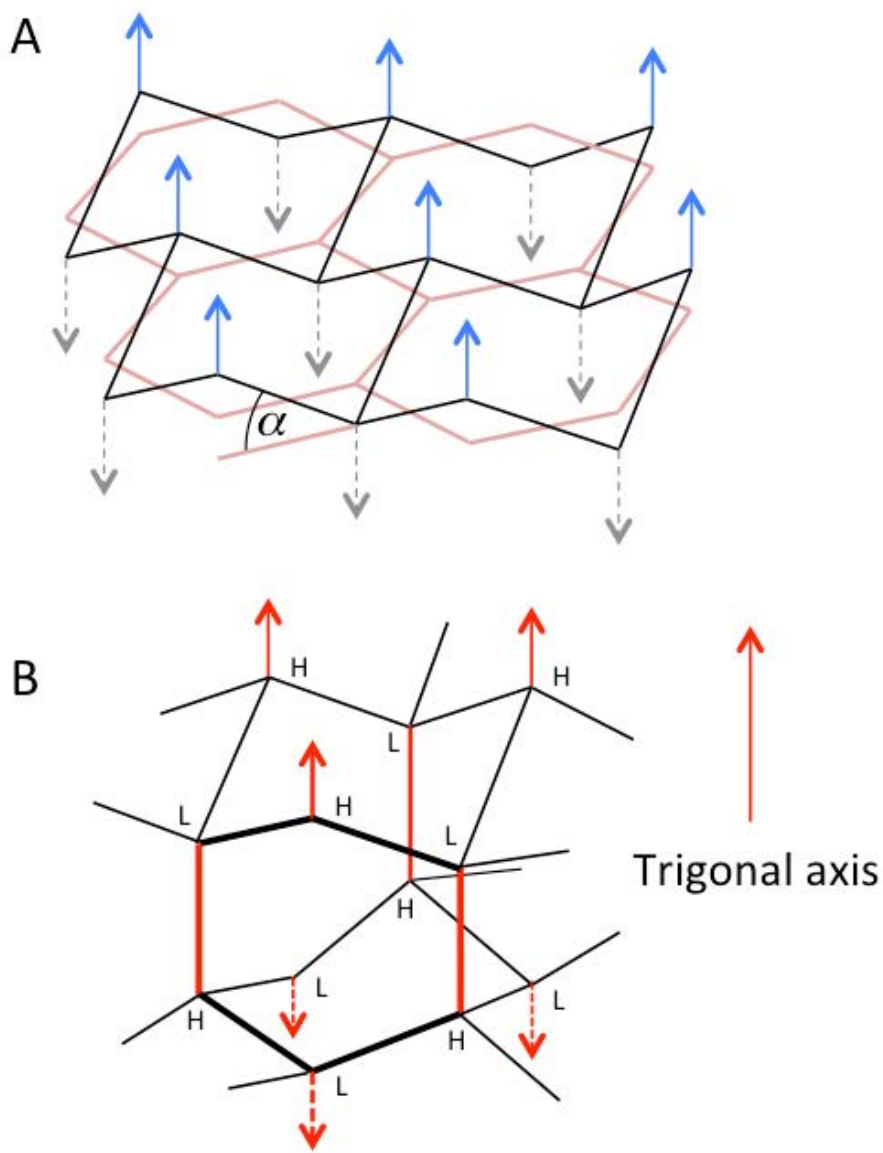

**Figure S2.4. Reproduced from [ref S2, Figure 10]. Part A: a plane lattice of hexagons (beige) is distorted to have alternating vertices above or below the plane, with the quasi-hexagon sides making angle $\alpha$ with the plane. Part B: two layers spaced along the vertical trigonal symmetry axis. High vertices in the lower layer connect to low vertices in the next layer up. Each connecting rod is identical.**

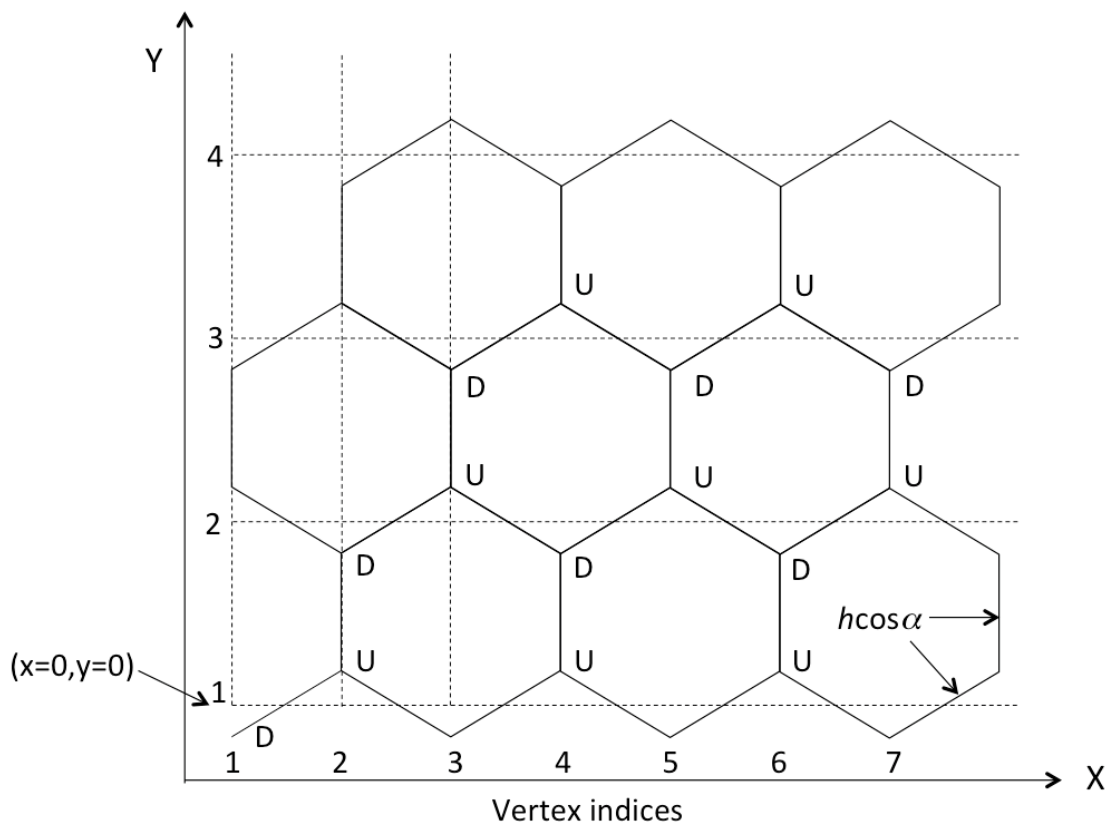

**Figure S2.5. View down the trigonal axis of the diamond 2H structure showing the structure's hexagonal projection with apparent length $h\cos\alpha$ of the connecting rods. Vertices are coded by X and Y numerals. Layers of the hexagonal projection are superimposed in a stack coming out of the page. From one layer to the next higher layer there are interchanged U and D characters that represent axial displacements of vertices upward and downward (along the z-axis). The layers are connected from a high point U (in a lower layer) to a low point D (in the layer above) by axial rods of length $h$. Each layer is coded by a Z numeral.**

Rather than seek a unit cell, which is conceptually difficult for this structure, we created (x,y,z) coordinates for each lattice vertex as a function of the numbers in an intuitive labeling system illustrated in Figure S2.5, which is a view down the trigonal symmetry axis. This takes as a starting point the hexagonal projection layers that lie in alignment vertically above and below each other to fill space, then adds or subtracts height from alternating vertices around the hexagons in the manner discussed in [2]. We need to have twice the number of "x" values as "y" and "z" values to uniquely specify the vertices of an approximately cubic lattice volume (Figure S2.5). In Figure S2.5 we are looking down upon fore-shortened sides to the hexagons, of apparent length $h\cos\alpha$, where $h$ is the true length of a connecting rod and $\alpha$ is the angle that each rod takes relative to the horizontal plane (illustrated in Fig. S2.4). When $\sin\alpha = 1/3$ the exact tetrahedral symmetry exists at every junction in the structure [2]. Here we allow angle $\alpha$ to be variable, giving access to the parameters of axially distorted structures.

Where *J,K,L* are the indices (coding numerals) along the *x,y,z* axes, the position of any vertex is given by

$$x = (J-1)\frac{\sqrt{3}}{2}h\cos\alpha$$

$$y = (K-1)\frac{3h}{2}\cos\alpha + (-1)^J \cdot (-1)^{(K-1)} \cdot \frac{h\cos\alpha}{4}$$

$$z = (L-1)(h + h\sin\alpha) + (-1)^J \cdot (-1)^{(K-1)} \cdot (-1)^{(L-1)} \cdot \frac{h\sin\alpha}{2}$$

Here for clarity in the superscripts we replaced the indices $J_X$, $J_Y$, $J_Z$ used in the program listed below by *J,K,L*.

With this apparatus we can consider index ranges suitable for a cube of side N quasi-cells (defined in [2]):

$1 < J < 2N$

$1 < K < N$

$1 < L < N$

A central point in the lattice is chosen, say at coordinates (N, N/2, N/2) with N even, then nearby lattice distances can be evaluated, for example in ranges for J , K , L

$N - 4 < J < N + 4$

$N/2 - 3 < K < N/2 + 3$

$N/2 - 3 < L < N/2 + 3$

When run with a maximum spacing cut-off of 140 Angstroms, this produces 69 spacings, many of which are duplicates. Further ordering and pruning to yield unique "*d*" values reduces this to a short list. In the comparison of data to theory that follows we chose a cutoff of 140Angstroms. When the exact tetrahedral angle is used, there are 9 distinct lattice "*d*" values less than 140 Angstroms. If an axial distortion is applied the degeneracy is broken and 12 values now appear under 140 Angstroms. Figure S2.6 plots the changes to these values as the deviation of $\alpha$ from tetrahedral ($\alpha$ = 19.471221 deg.) ranges through -5 to +5 degrees (the 49.0 Angstrom line, not shown, does not vary with $\alpha$).

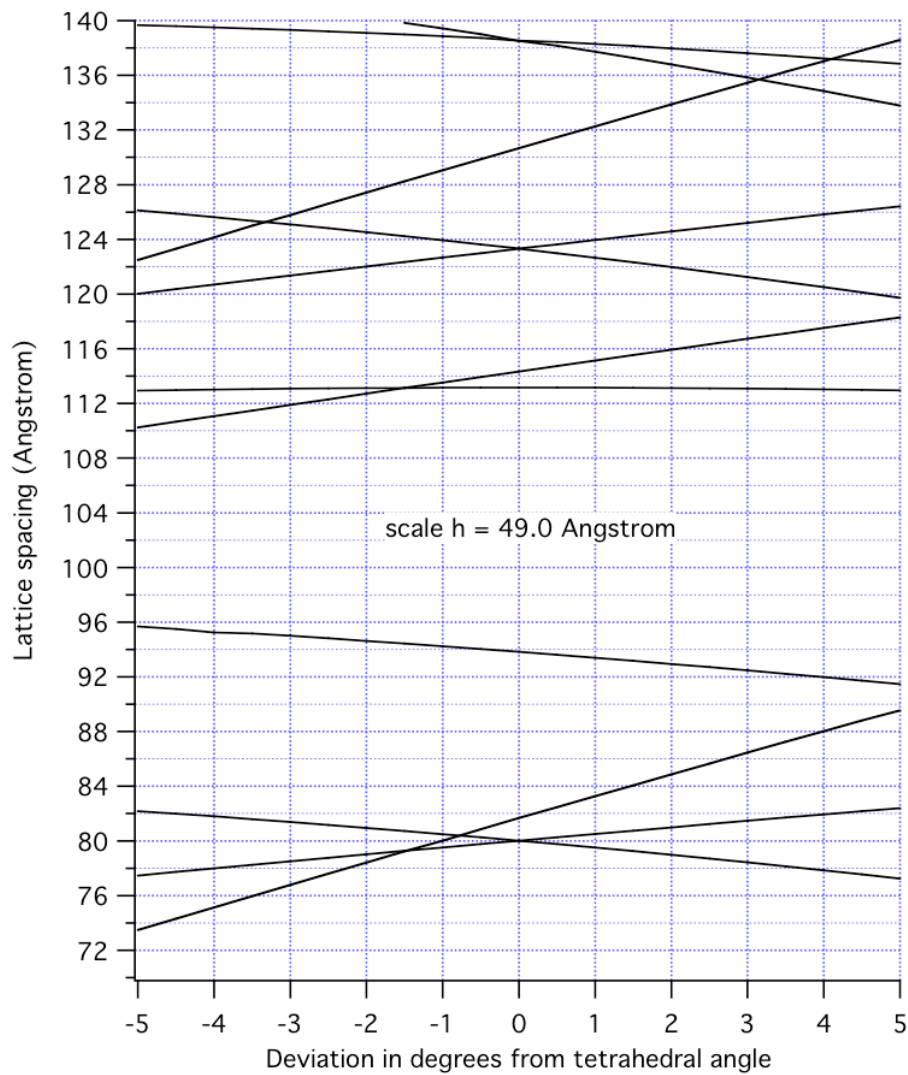

**Figure S2.6. Variation of lattice spacings under 140Angstrom when angle $\alpha$ is varied above and below the tetrahedral angle of 19.471221 degrees (relative to horizontal). The 49.0 Angstrom fundamental spacing does not vary with angle and is not shown on this plot. These results may be scaled to rod length X Angstroms by multiplying by X/49.0 .**

**Fit including lattice distortion to observed set**
It is found that a lattice spacing of 49.0 ±0.2 Angstroms combined with an angular deviation of +4.5 deg. gives the best fit to the reduced data of Figure S2.3 (fit shown as red bars). Only two major peaks in Figure S2.3 were not covered, at 98 and 130 Angstroms. Minor peaks were potentially due to random coincidences. The 98 Angstrom peak happens to be exactly two times the 49.0 Angstrom rod length. It can be produced automatically by doubling the diffraction orders {5,6,7 and 9} of column 2 in Table S2.2, therefore its presence is expected. The 130 Angstrom peak is not covered at +4.5 deg. in Figure S2.6, however it is covered precisely by the 0 deg. deviation non-distorted lattice. It could be that some parts of the lattice are un-distorted. The peak at 63 Angstrom is expected as a corollary of the 126.25

Angstrom peak which contains orders 14, 18, 20, 22, 24 that may be divided by 2 to give the valid orders 7, 9, 10, 11, 12, however no new lattice parameter is implied.

**S2 Discussion**

The use of wavelengths spanning the Fe K edge absorption allowed the iron-dependent aspects of ooid diffraction to stand out due to decreased absorption at the longer wavelength. It also appears to have been important to have reduced diffraction order numbers by use of a factor of two longer wavelength, contributing to a simpler analysis. It is considered, in conjunction with the ooid fluorescence data (main text), that the identified lattice is that of hemoglycin, which in prior x-ray work [2,6] has shown a "rod" length of 4.9nm. A clear analysis without competing structures suggests that this is the dominant and only iron-containing lattice within the ooid. A relatively small iron content of 0.15 wt % can produce this lattice, in conjunction with a glycine content of approximately 0.87 wt %. This considers a calcite-filled lattice with density 2.71. The density of aragonite varies from 2.93 to 2.95.

The lattice is filled with many calcium carbonate crystals, of a size that could be as small as the "cells" within the diamond 2H structure, i.e. between 5nm and 8nm across. Nano-crystals of nickel at this scale were observed in X-ray diffraction of a lattice sample from the Acfer 086 meteorite (analysis unpublished). Such fillings will vary depending upon the mix of atoms available as the lattice forms.

It is probable that a small degree of lattice distortion can be induced by any given crystal type within the cells. The cell volume is maximized for exact tetrahedral symmetry [2], but very little energy is required to induce an axial lattice distortion to accommodate a particular filling type of crystal. As the lattice is lengthened slightly along the trigonal axis the $h\cos\alpha$ projection decreases slightly, possibly providing an energy minimum for the combined system of lattice plus enclosed calcium carbonate crystals.

**Section S2 Appendix. Program to find diamond 2H lattice spacings**

```
'Programmed in qb64 for mac - qb64 is a generic Basic language
PRINT "Diamond Lattice"
PRINT "calculates x,y,z coordinates in a cube of side N"
PRINT "output is in a new file called dspace in the same folder as the program"
DIM Vx(22), Vy(11), Vz(11), set(15), fit(15), testall(100), testran(100), testwin(15)

INPUT "Variation (deg) from alpha at tetrahedral vertex ", dalpha
INPUT "number of quasi-cells (even and less than or equal to 10) ", N
INPUT "inter-vertex distance h (Angstroms) ", h

DIM dspace AS STRING
OPEN "dspace" FOR OUTPUT AS #1

pi = 3.14159265359
alp = (19.471221 + dalpha) * pi / 180
Salp = SIN(alp)
```

```
Calp = COS(alp)

Nx = 2 * N
Ny = N
Nz = N

Fx = SQR(3) * h * Calp / 2
Fy1 = 3 * h * Calp / 2
Fy2 = h * Calp / 4
Fz1 = h + h * Salp
Fz2 = h * Salp / 2

' set up central point N, N/2, N/2
N2 = N / 2

x0 = (N - 1) * Fx
y0 = (N2 - 1) * Fy1 + (-1) ^ N * (-1) ^ (N2 - 1) * Fy2
z0 = (N2 - 1) * Fz1 + (-1) ^ N * (-1) ^ (N2 - 1) * (-1) ^ (N2 - 1) * Fz2

'generate vertices around central point and distances from center

index = 1
FOR Jx = N - 4 TO N + 4
Vx(Jx) = (Jx - 1) * Fx

FOR Jy = N2 - 3 TO N2 + 3
Vy(Jy) = (Jy - 1) * Fy1 + (-1) ^ Jx * (-1) ^ (Jy - 1) * Fy2

FOR Jz = N2 - 3 TO N2 + 3
Vz(Jz) = (Jz - 1) * Fz1 + (-1) ^ Jx * (-1) ^ (Jy - 1) * (-1) ^ (Jz - 1) * Fz2

test = SQR((x0 - Vx(Jx)) ^ 2 + (y0 - Vy(Jy)) ^ 2 + (z0 - Vz(Jz)) ^ 2)
IF test > 140 THEN GOTO 16
testall(index) = test
index = index + 1

PRINT #1, USING "    .####^^^^"; Jx, Jy, Jz, test
16 'continue

NEXT Jz
NEXT Jy
NEXT Jx

PRINT "index = ", index
CLOSE #1
```

```
'rank list of "d" spacings in vector testall
rank = 1
FOR k = 1 TO index
trial = testall(k)

FOR m = 1 TO index
IF trial > testall(m) + 0.0001 THEN rank = rank + 1
NEXT m

testran(rank) = trial
rank = 1
NEXT k

'winnow testran to obtain unique set
q = 1
FOR p = 1 TO index
IF testran(p) > 0 THEN testwin(q) = testran(p) ELSE GOTO 20
q = q + 1
20 'continue
NEXT p

FOR g = 1 TO q - 1
PRINT testwin(g)
NEXT g
END
```

## Section S3: X-ray diffraction on crystals and ooids

### S3.1 Basic x-ray diffraction observations on fossil stromatolite and meteoritic crystals

In 2021 and 2022 at APS diffraction was studied in a range of crystals derived as described above from fossil stromatolite No. 1 Wyoming (2.1Gya), provenance in Section S 3.2.

In all except one crystal the dominant features were rings, and not spots, indicating generally multiple crystalline nature (Table S3.1). The stromatolite and Acfer 086 data is from beam line 31-1D-D, APS, at a wavelength of 0.9793 Angstroms. The vesicle hexagon data is from Diamond Light Source, at 1.000 Angstroms.

**Table S3.1: Rings measured on various crystals (Angstroms). All detectable rings are listed: gaps indicate no data. Sample to detector distances (mm) noted.**

| Stromatolite 272807 fossil No. 1 (400mm) | Stromatolite 275049 fossil No. 1 (329mm) | Stromatolite 289349 fossil No. 1 (190mm) | Rolled-up Hemoglycin sheet Acfer 086 [2] | Vesicle hexagon [9] Sutter'sMill |
|---|---|---|---|---|
| - | 6.28(5) ± 0.021 | 6.277 | 6.31 | - |
| 5.25 | 5.26 ± 0.016 | 5.251 | 5.29 | - |
| 4.70 | 4.78 ± 0.019 | 4.784 | 4.80 | - |
| - | 4.20 ± 0.011 | 4.196 | - | - |
| 4.03 | 4.05 ± 0.018 | 4.064 | - | 4.085 -only |
| 3.68 | 3.69 ± 0.013 | - | - | - |
| 3.45 | 3.49(5) ± 0.011 | 3.454 ± 0.043 | 3.52 | - |
| - | 3.34(5) ± 0.011 | 3.351 ± 0.034 | - | - |
| 3.06 | 3.09(3) ± 0.004 | 3.097 ± 0.033 | - | - |
| 2.89 | 2.89 ±0.007 | 2.892 ± 0.023 | - | - |
| 2.80 | - | - | - | - |
| 2.77 | - | - | - | - |
| 2.71 | 2.67(5) ± 0.005 | - | - | - |
| 2.62 | - | - | - | - |
| 2.51 | 2.54 ±0.007 | - | - | - |
| 2.40 | 2.40 ± 0.005 | - | - | - |
| 2.29 | - | - | - | - |
| 2.25 | - | - | - | |
| 2.19 | 2.195 ± 0.005 | - | - | |
| 2.14 | - | - | - | |
| 2.06 | - | - | - | |
| 2.01 | 2.015 ± 0.005 | - | - | |
| 1.94 | - | - | - | |
| 1.88 | - | - | - | |
| 1.84 | - | - | - | |
| 1.80 | 1.805 ± 0.005 | - | - | |

The first five entries in column 3 under 289349 were processed differently via scanning and subtraction of a relatively large background, so as to find accurate peak positions. The errors for these five points are systematic and of the order of 0.03 Angstroms. All other errors are one standard deviation of multiple ring measurements. Although the crystals in columns 2 and 3 do not appear to be externally similar at all, there is a very good lattice parameter match across all the major rings. Crystal 275049 was thin, purplish and well-shaped whereas crystal 289349 was a white "blob" with much greater depth. A high featureless background in the central part of the 289349 pattern could be due to amorphous material within its lattice. The first five entries under 289349 were, as a set, very much more intense than the subsequent entries.

We searched our meteorite and stromatolite data for matches to the above pair. In the 4th column we list the observed rings from a fiber crystal of Acfer 086 [2] the first two of which were identified with the Fe-Fe spacing at the four-way connection of hemoglycin rods. In the fifth column we list the single dominant spacing of Fe atoms at the three-way junction of hexagonal sheets of Sutter's Mill hemoglycin that formed a vesicle [9]. Although the latter may be a fortuitous match, these prior data resemblances to the stromatolite dominant rings could point to the presence of Fe-Fe spacings within fossil stromatolite similar to those of meteorite extract crystals.

**S 3.2 X-ray diffraction observations on present day and fossil ooids.**
At Diamond Light Source a series of x ray diffraction runs compared the crystal powder diffraction patterns of **ooids** from the following stromatolite sources:

**Sample Sources:** Present-day stromatolites are supplied by Andrew Knoll of the Museum of Comparative Zoology and Organismic and Evolutionary Biology (OEB) Harvard.
Present-day stromatolite details are:
3. Shark-Bay Western Australia – collected by Elso Barghoorn 1971- estimated to be 2000-3000 years old.
4. **K-05 SS-1** from San Salvador Island Bahamas – collected by Andrew Knoll in 2005 –- A modern mineralized microbialite.

Fossil stromatolite details are:
3. 2.1Ga stromatolite No. 1 from Medicine Bow region, Wyoming – collected by David Lageson of Montana State University.
4. 2.1Ga stromatolite No. 2 from Medicine Bow region, Wyoming – collected by David Lageson of Montana State University.

X-ray diffraction results for ooids in the two present day stromatolites and in a second Wyoming sample, No. 2, are compared in Table S3.2. No ooids could be found in Fossil stromatolite No. 1.

**Table S3.2 Comparison of present day and fossil ooid diffraction patterns on Diamond Light Source. All mounted samples are multiple crystals, giving rings. The number of x ray runs that were averaged is given. Almost uniformly the rings identify as aragonite, with one possible calcite ring.**

| Ooids, x ray 1.000A, all data in cryo-stream at 100K, "d" in Angstroms ||||| 
|---|---|---|---|---|
| Shark Bay No acid n = 2 | Shark Bay Acid treated n = 2 | 2nd Fossil Stromatolite n = 3 | San Salvador n = 1 | Assignment *a* = aragonite *c* = calcite |
|  | 16.605 |  |  |  |
|  | 11.65 |  |  |  |
|  |  | 10.458 |  |  |
| 10.325 |  |  | 10.32 |  |
|  | 8.605 |  |  |  |
|  | 8.246 |  |  |  |
|  | 7.424 |  |  |  |
|  |  | 7.11 | 7.093 |  |
|  | 6.905 |  |  |  |
|  | 6.376 |  |  |  |
|  | 5.505 |  |  |  |
|  |  | 4.38 |  |  |
| 4.194 |  | 4.194 |  |  |
| 3.791 | 3.819 | 3.812 |  |  |
|  | 3.493 |  |  |  |
| 3.388 | 3.393 | 3.387 |  | a |
| 3.263 | 3.263 | 3.262 | 3.265 | a |
|  | 3.213 |  |  |  |
| 2.988 | 2.988 | 2.993 | 3.00 |  |
| 2.860 | 2.865 |  |  |  |
| 2.725 |  | 2.70 |  | a |
| 2.690 |  | 2.689 | 2.698 | a |
| 2.480 | 2.47 | 2.472 | 2.475 | a |
| 2.40 |  | 2.407 |  |  |
| 2.363 | 2.358 | 2.367 | 2.37 | a |
| 2.328 | 2.328 | 2.324 | 2.325 | a |
| 2.248 | 2.245 |  |  |  |
| 2.185 |  | 2.182 |  |  |
| 2.065 |  | 2.08 |  |  |
| 1.978 | 1.968 | 1.974 | 1.975 | a |
| 1.877 | 1.875 | 1.876 | 1.878 | c , a |
| 1.808 | 1.808 | 1.807 | 1.81 |  |
|  | 1.738 | 1.737 | 1.74 | a |
|  |  | 1.719 | 1.72 |  |
| 1.60 |  |  |  |  |